\documentclass[reprint, superscriptaddress, amsmath,amssymb, aps, prl]{revtex4-1}

\newcommand{\lp}{\left(}
\newcommand{\rp}{\right)}
\newcommand{\lab}{\left<}
\newcommand{\rab}{\right>}
\newcommand{\lsb}{\left[}
\newcommand{\rsb}{\right]}

\newcommand{\labs}{\left|}
\newcommand{\rabs}{\right|}

\newcommand{\grad}{\nabla}
\newcommand{\tor}{\Phi}

\newcommand{\rudolphpeierls}{Rudolph Peierls Centre for Theoretical Physics, University of Oxford, 1 Keble Road,  Oxford, OX1 3NP, UK}
\newcommand{\blackett}{Blackett Laboratory, Imperial College, London, SW7 2AZ, UK}
\newcommand{\culham}{EURATOM/CCFE Fusion Association, Culham Science Centre, Abingdon, OX14 3DB, UK}

\newcommand{\mertoncollege}{Merton College, Oxford, OX1 4JD, UK}
\newcommand{\mitpsfc}{Plasma Science and Fusion Center, MIT, Cambridge, MA 02139, USA}

\newcommand{\kfki}{Wigner Research Centre for Physics, Association EURATOM/HAS, P.O. Box 49, H-1525, Budapest, Hungary}
\newcommand{\orise}{Oak Ridge Institute for Science and Education, Oak Ridge, TN 37831, USA}

\newcommand{\ycghim}{\author{Y.-c.~Ghim}\email{y.kim1@physics.ox.ac.uk}\affiliation{\rudolphpeierls}\affiliation{\culham}}
\newcommand{\aaschekochihin}{\author{A.~A.~Schekochihin}\affiliation{\rudolphpeierls}\affiliation{\mertoncollege}}
\newcommand{\arfield}{\author{A.~R.~Field}\affiliation{\culham}}
\newcommand{\scowley}{\author{S.~C.~Cowley}\affiliation{\culham}\affiliation{\blackett}}

\newcommand{\ddunai}{\author{D.~Dunai}\affiliation{\kfki}}
\newcommand{\szoletnik}{\author{S.~Zoletnik}\affiliation{\kfki}}
\newcommand{\mastteam}{\author{the~MAST~Team}\affiliation{\culham}}

\newcommand{\fparra}{\author{F.~I.~Parra}\affiliation{\mitpsfc}}
\newcommand{\mbarnes}{\author{M.~Barnes}\affiliation{\mitpsfc}\affiliation{\orise}}
\newcommand{\gcolyer}{\author{G.~Colyer}\affiliation{\rudolphpeierls}\affiliation{\culham}}
\newcommand{\iabel}{\author{I.~G.~Abel}\affiliation{\rudolphpeierls}\affiliation{\mertoncollege}}

\usepackage{graphicx}
\usepackage{dcolumn}
\usepackage{bm}
\usepackage{color}

\newcommand{\vti}{\ensuremath{v_{{\rm th}i}}}

\newcommand{\vtie}{\ensuremath{v_{{\rm th}i, e}}}
\newcommand{\rhoi}{\rho_i}

\newcommand{\rhoie}{\rho_{i, e}}
\newcommand{\lx}{\ell_x}
\newcommand{\ly}{\ell_y}

\newcommand{\lpar}{\ell_\parallel}

\newcommand{\lZ}{\ell_Z}

\newcommand{\tc}{\tau_{\rm c}}

\newcommand{\tnlnz}{\tau_{\rm nl}^{\rm NZ}}
\newcommand{\tst}{\tau_{\rm st}}
\newcommand{\tshear}{\tau_{\rm sh}}
\newcommand{\tM}{\tau_{\rm M}}
\newcommand{\tstar}{\tau_\ast}
\newcommand{\tstari}{\tau_{\ast i}}
\newcommand{\tstare}{\tau_{\ast e}}
\newcommand{\tstarie}{\tau_{\ast i, e}}
\newcommand{\tstarn}{\tau_{\ast n}}

\newcommand{\delvper}{\delta u_\perp}
\newcommand{\dn}{\delta n}
\newcommand{\dI}{\delta I}
\newcommand{\LTi}{L_{T_i}}

\newcommand{\LTie}{L_{T_{i, e}}}
\newcommand{\Ln}{L_n}

\newcommand{\Lstar}{L_*}
\newcommand{\corr}{\mathcal{C}}
\newcommand{\Dx}{\Delta x}

\newcommand{\DZ}{\Delta Z}
\newcommand{\Dt}{\Delta t}

\newcommand{\vbes}{v_{BES}}
\newcommand{\taupeak}{\Dt_{\rm peak}}
\renewcommand{\eqref}[1]{Eq. (\ref{#1})}
\newcommand{\myfig}[3][5in]{
\begin{figure}[tbp]
\includegraphics[width=#1]{#2}%
\caption{#3\label{fig:#2}}%
\end{figure}
}

\newcommand{\figref}[1]{Fig.~\ref{fig:#1}}

\begin{document}


\title{Experimental Signatures of Critically Balanced Turbulence in MAST}
\ycghim
\aaschekochihin
\arfield
\iabel
\mbarnes
\gcolyer
\scowley
\fparra
\ddunai
\szoletnik
\mastteam

\date{\today}

\begin{abstract}
Beam Emission Spectroscopy (BES) measurements of ion-scale density fluctuations in the MAST tokamak are used to show that the turbulence correlation time, the drift time associated with ion temperature or density gradients, the particle (ion) streaming time along the magnetic field and the magnetic drift time are consistently comparable, suggesting a ``critically balanced'' turbulence determined by the local equilibrium. The resulting scalings of the poloidal and radial correlation lengths are derived and tested. The nonlinear time inferred from the density fluctuations is longer than the other times; its ratio to the correlation time  scales as $\nu_{*i}^{-0.8\pm0.1}$, where $\nu_{*i}=$ ion collision rate/streaming rate. This is consistent with turbulent decorrelation being controlled by a zonal component, invisible to the BES, with an amplitude exceeding the drift waves' by~$\sim \nu_{*i}^{-0.8}$. 
\end{abstract}

\maketitle

\paragraph{Introduction.} 
Microscale turbulence hindering energy confinement in magnetically confined hot plasmas is driven by gradients of equilibrium quantities such as temperature and density. These gradients give rise to instabilities that inject energy into plasma fluctuations (``drift waves'') at scales just above the ion Larmor scale. The most effective of these is believed to be the ion-temperature-gradient (ITG) instability \cite{rudakov_doklady_1961,coppi_pof_1967,cowley_pfb_1991}. A turbulent state ensues, giving rise to ``anomalous transport'' of energy \cite{horton_rmp_1999}. It is of interest, both for practical considerations of improving confinement and for the fundamental understanding of multiscale plasma dynamics, what the structure of this turbulence is and how its amplitude, scale(s) and resulting transport depend on the equilibrium parameters: ion and electron temperatures, density, angular velocity, magnetic geometry, etc. 
\newline\indent
Fluctuations in a magnetized toroidal plasma are subject to a number of distinct physical effects, which can be thought about in terms of various time scales such as the drift times associated with the temperature and density gradients, the particle streaming time along the magnetic field as it takes them around the torus toroidally and poloidally, the magnetic ($\nabla B$ and curvature) drift times of particles moving across the field, the nonlinear time of the fluctuations being advected across the field by the fluctuating $\vec{E}\times\vec{B}$ velocity, the time between collisions, the shear time associated with plasma rotation. Some of these time scales and, consequently, the corresponding physics may be irrelevant, while others play a crucial role for the saturation of the linearly unstable fluctuations. There has been a growing understanding \cite{barnes_prl_2011_107}, driven largely by theory \cite{goldreich_apj_1995,cho_apj_2004,schekochihin_apjs_2009,nazarenko_jfm_2011},  observations \cite{horbury_prl_2008,podesta_apj_2009,wicks_mnras_2010} and simulations of magnetohydrodynamic \cite{cho_apj_2000,maron_apj_2001,chen_mnras_2011} and kinetic \cite{cho_apj_2004,tenbarge_pop_2012} plasma turbulence in space, that if a medium can support parallel (to the magnetic field) propagation of waves (and/or particles) and nonlinear interactions in the perpendicular direction, the turbulence in such a medium would normally be ``critically balanced,'' meaning that the characteristic time scales of propagation and nonlinear interaction would be comparable to each other and (therefore) to the correlation time of the fluctuations. This means that the turbulence is {\em not} weak and {\em not} two-dimensional, unless specially constrained to be so \cite{nazarenko_jfm_2011}. 
\newline\indent
Beam Emission Spectroscopy (BES) measurements of density fluctuations in tokamak plasmas \cite{fonck_rsi_1990,fonck_prl_1993,mckee_rsi_1999,mckee_rsi_2003,field_rsi_2012} have made it possible to probe ion-scale turbulence in these devices directly. In this Letter, we use such measurements in the MAST tokamak, along with the local equilibrium parameters calculated by other diagnostics, to estimate and compare the characteristic time scales of the turbulent fluctuations in the energy-containing range. We obtain, for the first time, direct evidence that the correlation, drift and parallel streaming time scales are indeed comparable across a range of equilibrium parameters (cf. \cite{mckee_nf_2001, hennequin_ppcf_2004}) and that the magnetic drift time is part of this ``grand critical balance'' as well. We also find indirect evidence that the decorrelation rate of turbulence is controlled by a zonal component whose relative importance to the drift-wave-like fluctuations scales with the ion collisionality.  
\newline\indent
Before presenting this evidence and its implications (e.g., dependence of the  correlation lengths on equilibrium parameters), let us describe how it was obtained.    
 
\paragraph{Experimental data and its analysis.} 
\myfig[3in]{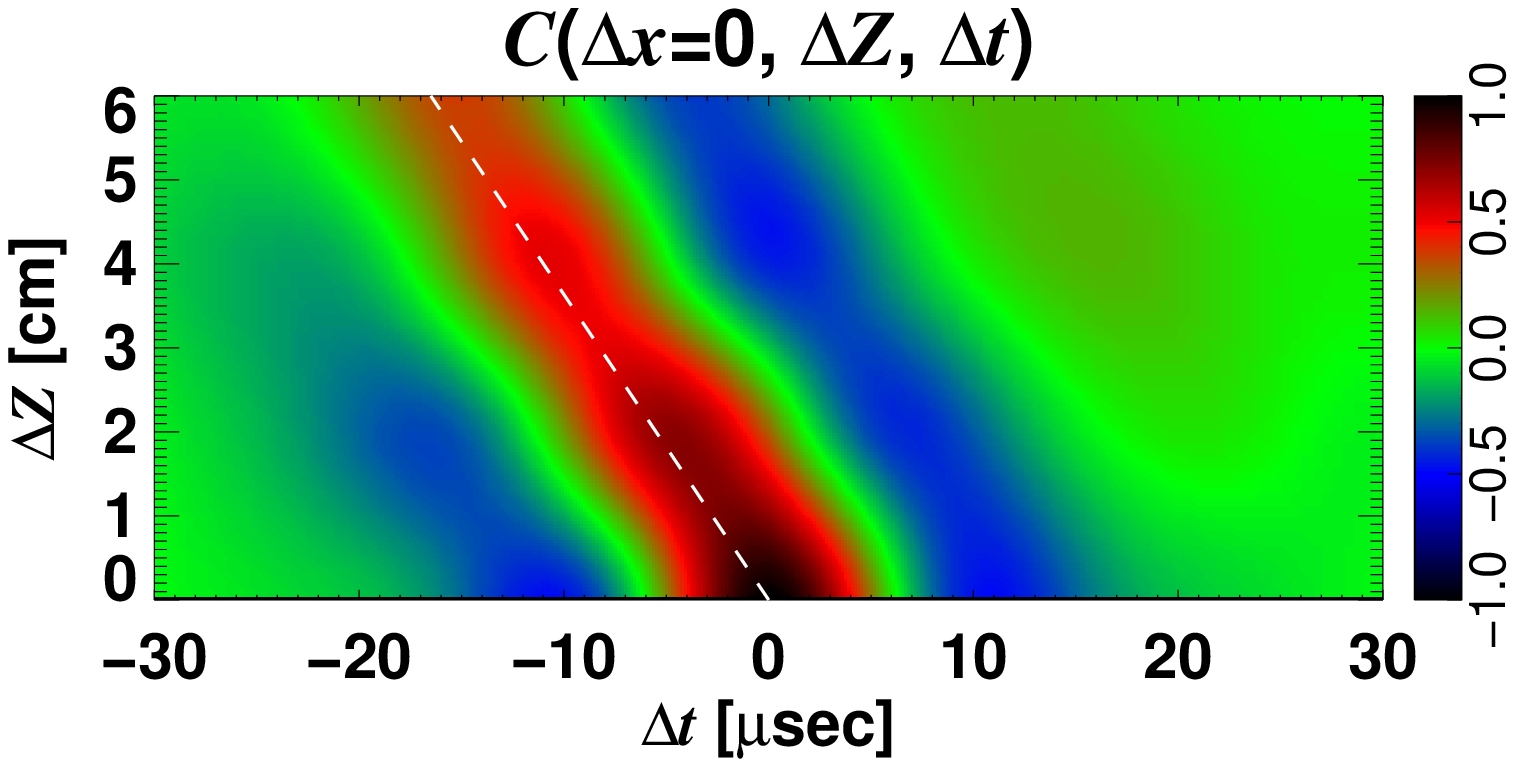}{An example of the correlation function in the poloidal-temporal plane, $\corr\lp\Dx=0,\DZ,\Dt\rp$. This data was taken at $r=30$\:cm, toroidal rotation speed was $U_\phi=10$\:km/s and magnetic pitch angle $\alpha=20^\circ$. The direction of maximum correlation is the direction of the magnetic field (dashed line).}
During the 2011 campaign, density fluctuation data from the BES diagnostic \cite{field_rsi_2012} on MAST were collected in a variety of discharges (including L- and H-modes and internal transport barriers). Here we report the data from 39 neutral-beam heated ``double-null-diverted'' discharges, with no pellet injection and no resonant magnetic perturbations. The BES system on MAST collects photons from a 2D array of 8 radial $\times$ 4 vertical locations in the outboard midplane of the tokamak, with 2\:cm separation between the adjacent channels in either direction. The detected photon intensity (mean $+$ fluctuating, $I+\dI$) is used to infer, at each location, the density fluctuation level $\dn/n=\lp1/\beta\rp\lp\dI/I\rp$ \cite{fonck_rsi_1990}, where $\beta$ depends on the mean density $n$ and is estimated based on the Hutchinson model \cite{hutchinson_ppcf_2002} (dependence on the mean temperature is weak). As the BES array was moved radially for different discharges, our database contains cases with radial viewing positions $10$\:cm\:$<r<50$\:cm from the magnetic axis (the minor radius of the plasma is $\approx60$\:cm). 
\newline\indent
Local equilibrium parameters are measured by standard diagnostics: mean electron densities $n_e$ and temperatures $T_e$ by the Thomson scattering system \cite{scannell_rsi_2010}, impurity ion (C$^{6+}$) mean temperatures (assumed to equal the bulk ion temperature $T_i$) and toroidal flow velocity $U_\phi$ by the Charge eXchange Recombination Spectroscopy (CXRS) system \cite{conway_rsi_2006}, local magnetic pitch angle $\alpha$ by the Motional Stark Effect (MSE) system \cite{debock_rsi_2008}, and further equilibrium magnetic field information is obtained from pressure- and MSE-constrained \texttt{EFIT} equilibria \cite{lao_nf_1985}.
\newline\indent
We filter the BES data to the frequency interval $[20,100]$\:kHz and calculate the spatio-temporal correlation function
\begin{align}
\label{eq:corr_def}
\nonumber
&\corr\lp\Dx,\DZ,\Dt\rp = \\
&\frac{\lab \dI\lp x, Z, t \rp \dI\lp x+\Dx, Z+\DZ, t+\Dt\rp \rab}{\sqrt{\lab\dI^2\lp x, Z, t\rp\rab \lab\dI^2\lp x+\Dx, Z+\DZ,  t+\Dt\rp\rab}},
\end{align}
where $x$, $Z$ and $t$ are the radial, vertical and time coordinates, respectively, and $\Dx$, $\DZ$ and $\Dt$ are the corresponding channel separations and the time lag; $\lab \cdot \rab$ is the time average over 5\:ms periods. At $\Dx=\DZ=0$, the auto-covariances $\lab \dI\lp x, Z, t \rp \dI\lp x, Z, t+\Dt\rp \rab$ contain not only the physical signal but also photon and electronic noise. We remove this effect by applying LED light to the BES channels, obtaining 150 different DC levels of BES signal from $0$ to $1.5$\:V, calculating the noise auto-covariance $C_N\lp\Dt\rp$ at each DC level with the same band frequency filter of $[20,100]$\:kHz, then finding $C_N\lp\Dt\rp$ whose DC level of the signal matches the DC level of the BES data from the MAST discharges, and subtracting it from the calculated auto-covariances. From the correlation function (\ref{eq:corr_def}) (illustrated in \figref{spatio_temporal_cont1}), we calculate the local characteristics of the density fluctuations. 
\newline\indent
The fluctuation level at each radial location is obtained from the (noise-subtracted) auto-covariance function $\dn/n = \lp1/\beta\rp\sqrt{\lab\dI^2(x,Z,t)\rab}/I$ at all 32 locations and then averaged over the four poloidally separated channels at the same radial location. 
\newline\indent
The correlation length $\ly$ in the direction parallel to the flux surface and perpendicular to the magnetic field is obtained from the vertical (poloidal) correlation length $\lZ$ via $\ly=\lZ\cos\alpha$, assuming that the parallel correlation length is sufficiently long: $\lpar\gg\ly\tan\alpha$. The correlation length $\lZ$ is estimated using four poloidal channels at each radial location (the top channel is the reference channel) by fitting $\corr\lp\Dx=0, \DZ, \Dt=0\rp$ to the function $f_Z\lp\DZ\rp = p_Z + \lp 1 -p_Z \rp\cos\lsb2\pi \DZ/\lZ\rsb\exp\lsb-\labs\DZ\rabs/\lZ\rsb$, where $p_Z$ is a fitting constant that serves to account for global structures such as coherent MHD modes (for which $\corr\lp \Dx=0, \DZ=\infty, \Dt=0\rp = p_Z \neq 0$). In choosing $f_Z\lp\DZ\rp$, we assumed wave-like fluctuations in the poloidal direction \cite{fonck_prl_1993} (drift-wave turbulence), with the wavelength and correlation length comparable to each other. It is not possible to distinguish meaningfully between the two with only four poloidal channels. Assuming wave-like structure is essential as in most cases, we find that $\corr\lp\Dx=0, \DZ, \Dt=0\rp$ goes negative and/or is non-monotonic over the vertical extent of the BES array. 
\newline\indent
The radial correlation length $\lx$ is estimated using eight radial channels at each poloidal location (the fourth channel from the inward side is the reference channel). The correlation function $\corr\lp\Dx, \DZ=0, \Dt=0\rp$ is fitted to the function $f_x\lp\Dx\rp = p_x + \lp 1 -p_x \rp\exp\lsb-\labs\Dx\rabs/\lx\rsb$, where $p_x$ plays the same role as $p_Z$ did for $f_Z$. The values of $\lx$ from four poloidal locations are averaged, assuming that the radial correlations do not change significantly within the poloidal extent of the BES array. Because we have to use the entire array to estimate $\lx$, the number of data points for $\lx$ is 8 times smaller than for $\ly$.
\newline\indent
To estimate the correlation time $\tc$, we use the fact that the fluctuating density patterns are advected poloidally past the BES array with an apparent velocity $\vbes = U_\phi\tan \alpha$ due to the toroidal rotation velocity $U_\phi$ \cite{ghim_ppcf_2012}. We fit $\corr\lp\Dx=0, \DZ, \Dt=\taupeak\lp\DZ\rp\rp$ taken at the time delay $\taupeak\lp\DZ\rp$ when the correlation function is maximum at a given $\DZ$ \cite{durst_rsi_1992}, to the function $f_\tau\lp\DZ\rp = \exp\lsb -\labs\taupeak\lp\DZ\rp\rabs/\tc\rsb$. The reliability of this method relies on the temporal decorrelation dominating over the parallel spatial decorrelation, viz., we require $\tc\ll\lpar\cos\alpha/U_\phi$. Anticipating the critical balance assumption $\tc\sim \lpar/\vti$ \cite{barnes_prl_2011_107}, where $\vti = \sqrt{2T_i/m_i}$ is the ion thermal speed, and denoting the Mach number ${\rm Ma}=U_\phi/\vti$, we estimate that the fractional error in $\tc$ is $\sim {\rm Ma}/\cos\alpha$, which was never more than 20\% in the MAST discharges we used.
\newline\indent
The four quantities $\dn/n$, $\ly$, $\lx$ and $\tc$ are calculated at $8$ radial locations (except $\lx$), every $5$\:ms for all 39 discharges. All the fits described above are obtained via the \texttt{mpfit} procedure \cite{markwardt_mpfit}. We consider a data point unreliable and remove it from the database if (i) $I<0.3$\:V (the signal-to-noise ratio is too low); (ii) the estimated correlation lengths are smaller than the distance between the channels, $\lx$ or $\ly < 2$\:cm; (iii) the assumption that plasma rotation is mostly toroidal is suspect, viz., $\labs\lp\vbes-U_\tor\tan\alpha\rp/\vbes\rabs \ge 0.2$ (see ref. \cite{ghim_ppcf_2012}), where $\vbes$ is calculated at each radial location using the cross-correlation time delay (CCTD) method \cite{durst_rsi_1992}; (iv) the estimated error in the calculation of  $\vbes$ is $>20\%$; (v) $p_Z$ or $p_x >0.5$. The last two exclusion criteria pick out the cases when MHD modes are too strong; they are known to degrade the reliability of the BES data \cite{ghim_ppcf_2012}. The remaining database contains 448 points. 

\paragraph{Correlation time vs.\ drift time.} 
The turbulence can be driven by radial gradients in the mean ion and electron temperatures $T_{i,e}$ and density $n$. Denoting $\LTie^{-1}=|\grad\ln T_{i, e}|$ and $\Ln^{-1}=|\grad\ln n|$, the associated time scales are the inverse drift frequencies: 
\begin{equation}
\tstarie^{-1} = \frac{\rhoie}{\ly}\frac{\vtie}{\LTie},\quad 
\tstarn^{-1} = \frac{\rhoi}{\ly}\frac{\vti}{\Ln}, 
\end{equation}
where $\rhoie=\vtie/\Omega_{i,e}$ are the ion ($i$) and electron ($e$) Larmor radii, $\vtie=\sqrt{2T_{i, e}/m_{i, e}}$ the thermal speeds and $\Omega_{i,e}=eB/m_{i,e}c$ the Larmor frequencies. To estimate the drift times, we need information about the local equilibrium ($T_{i,e}$, $\LTie$ $\Ln$, $B$) and the correlation length $\ly$, calculated from the poloidal BES correlations. 
\newline\indent
\myfig[3in]{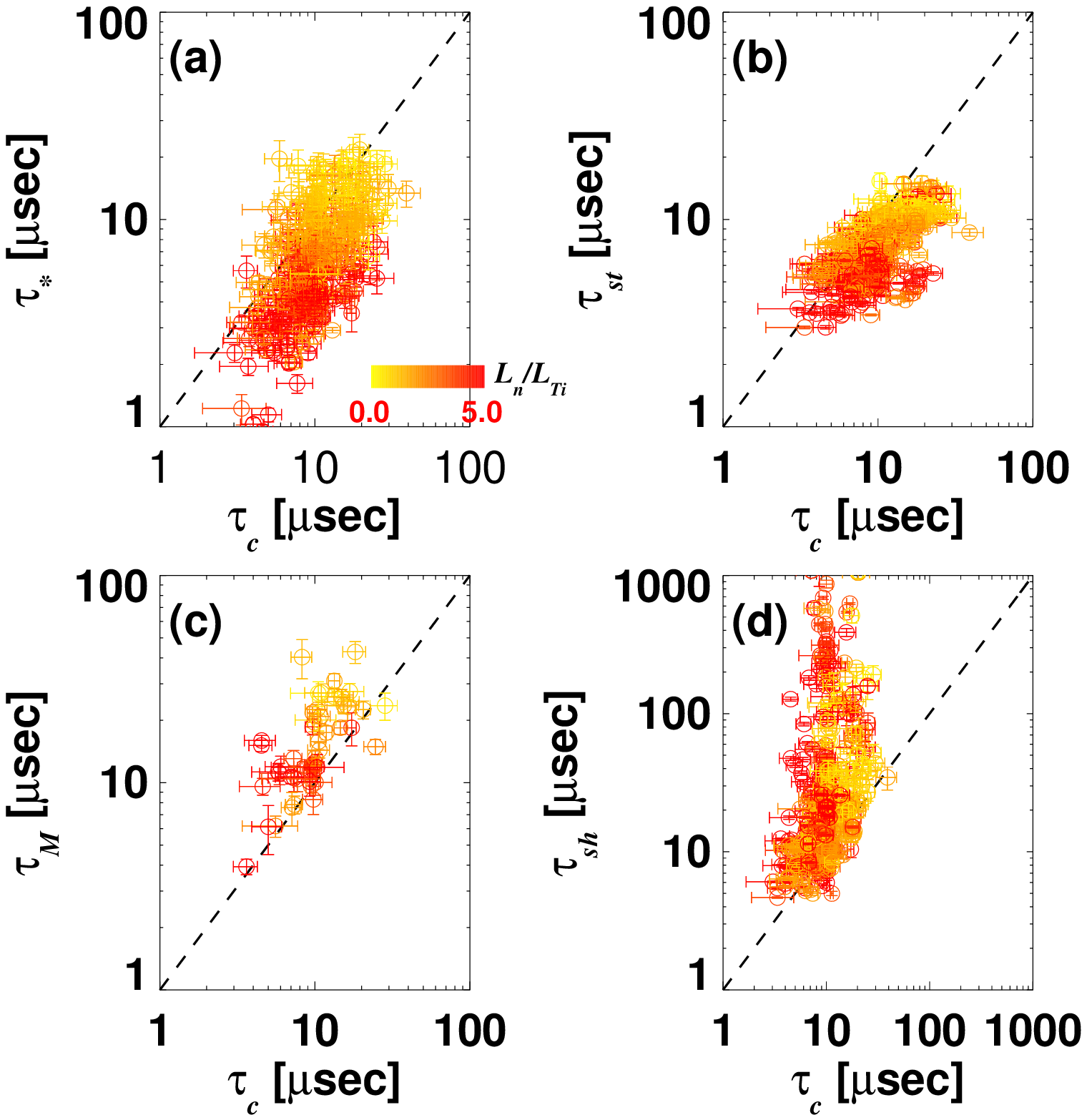}{(a) Drift time $\tstar=(\ly/\rhoi)\Lstar/\vti$ vs.\ correlation time $\tc$; (b) streaming time $\tst=\Lambda/\vti=(B/B_p)\pi r/\vti$ vs.\ $\tc$; (c) magnetic drift time $\tM = (\lx/\rhoi)R/\vti$ vs.\ $\tc$; (d) perpendicular velocity shear time $\tshear=[(B_p/B)dU_\phi/dr]^{-1}$ vs.\ $\tc$. In all cases, the color of points represents $\eta_i=\Ln/\LTi$.}
In \figref{all_vs_tau_c_eta}(a), we compare the drift times with the correlation time $\tc$ calculated from the spatio-temporal BES correlations. We find that $\tstar=(0.7\pm0.3)\tc$, where $\tstar=\min\{\tstari,\tstarn\}$ and the spread is calculated as the root mean square deviation from the mean value. The scaling holds over an order of magnitude in either time scale. Thus, the turbulence appears to be driven by the larger of the ion temperature or density gradient \footnote{However, for $\tc\lesssim10\:\mu$sec, $\tstare\sim\tstari$ and for $\tc\gtrsim10\:\mu$sec, $\tstarn\sim\tstari$, so we cannot rule out ion-scale electron drive (e.g., trapped electron modes \cite{kadomtsev_nf_1971} or microtearing \cite{roach_ppcf_2005,guttenfelder_pop_2012,doerk_pop_2012}).}. We find no clear correlation of $\tstare$ with $\tc$, or with any of the other time scales discussed below. 

\paragraph{Critical balance.} 
The standard argument behind the critical balance conjecture is causality \cite{nazarenko_jfm_2011}: two distant points on a field line cannot stay correlated if information cannot be exchanged between them over a turbulence correlation time. Assuming information travels at $\vti$, one gets $\lpar\sim\vti\tc$. This cannot be checked directly because there are no diagnostics capable of measuring $\lpar$ on MAST \footnote{As noted above, our method for measuring $\tc$ would instead yield $\lpar/\vti$ if ${\rm Ma}>\cos\alpha$, but that would require much stronger rotation (the smallest value in our database is $\cos\alpha\approx0.76$).}. Considering that the inboard side of the torus is a region of ``good'' (stabilizing) curvature, not much turbulence is expected there, so we assume that, at the energy injection scale, $\lpar\sim\Lambda$ \cite{barnes_prl_2011_107}, where the distance along the field line that takes a particle from the outer to the inner side of the torus is $\Lambda=\pi r B/B_p$ ($r$ is the minor radius at the BES position on the outer side and $B_p$ the poloidal component of the magnetic field) \footnote{In a conventional tokamak, $\Lambda\approx \pi q R$, where $q$ is the safety factor and $R$ major radius, but in a spherical tokamak, the local estimate we use is more appropriate}. Then critical balance means that $\tc$ should be comparable to 
\begin{equation}
\tst^{-1}=\frac{\vti}{\Lambda} = \frac{\vti}{\pi r}\frac{B_p}{B}\sim\frac{\vti}{\lpar},
\end{equation}  
the ion streaming time 
(the first two equalities are its definition, the last an assumption). Indeed, we find $\tst=(0.8\pm0.3)\tc$ (see \figref{all_vs_tau_c_eta}(b)). 
\newline\indent
The balance $\tst\sim\tstar$ implies that the poloidal correlation scale is $\ly/\rhoi\sim \Lambda/\Lstar$, where $\Lstar=\min\{\LTi,\Ln\}$ \cite{barnes_prl_2011_107}. This is tested in \figref{spatial_eta}(a), showing that while the two quantities are certainly of the same order, we do not have enough of a range of equilibrium parameters to state conclusively that this theoretically predicted scaling works.  
\paragraph{Magnetic drift time and radial correlation scale.} 
The time scale of the magnetic ($\nabla B$ and curvature) drifts is
\begin{equation}
\tM^{-1}=\frac{\rhoi}{\lx}\frac{\vti}{R},
\end{equation}
where we have assumed that the scale length of the background magnetic field is $R$ (major radius at the viewing location) and $\lx<\ly$ (this will shortly prove correct). It is clear that this scale cannot be shorter than $\tc$ because damping due to the drift resonance would eliminate such fluctuations. While magnetic drift physics may matter (in a torus, curvature contributes to the ITG drive \cite{horton_rmp_1999}), it does not have to affect scalings, as, for example, it would not in a slab and as it did not in the numerical simulations of \cite{barnes_prl_2011_107}. In contrast, \figref{all_vs_tau_c_eta}(c) shows that in the MAST discharges we have analyzed, $\tM$ is not negligible and scales with $\tc$, similarly to $\tstar$ and $\tst$ \footnote{As $\tM$ contains $\lx$, there are 8 times fewer data points here than in previous two figures, as explained above}. We find $\tM=(1.6\pm0.7)\tc$. Thus, a ``grand critical balance'' appears to hold in MAST, viz., $\tc\sim\tstar\sim\tst\sim\tM$. 
\newline\indent
\myfig[3.0in]{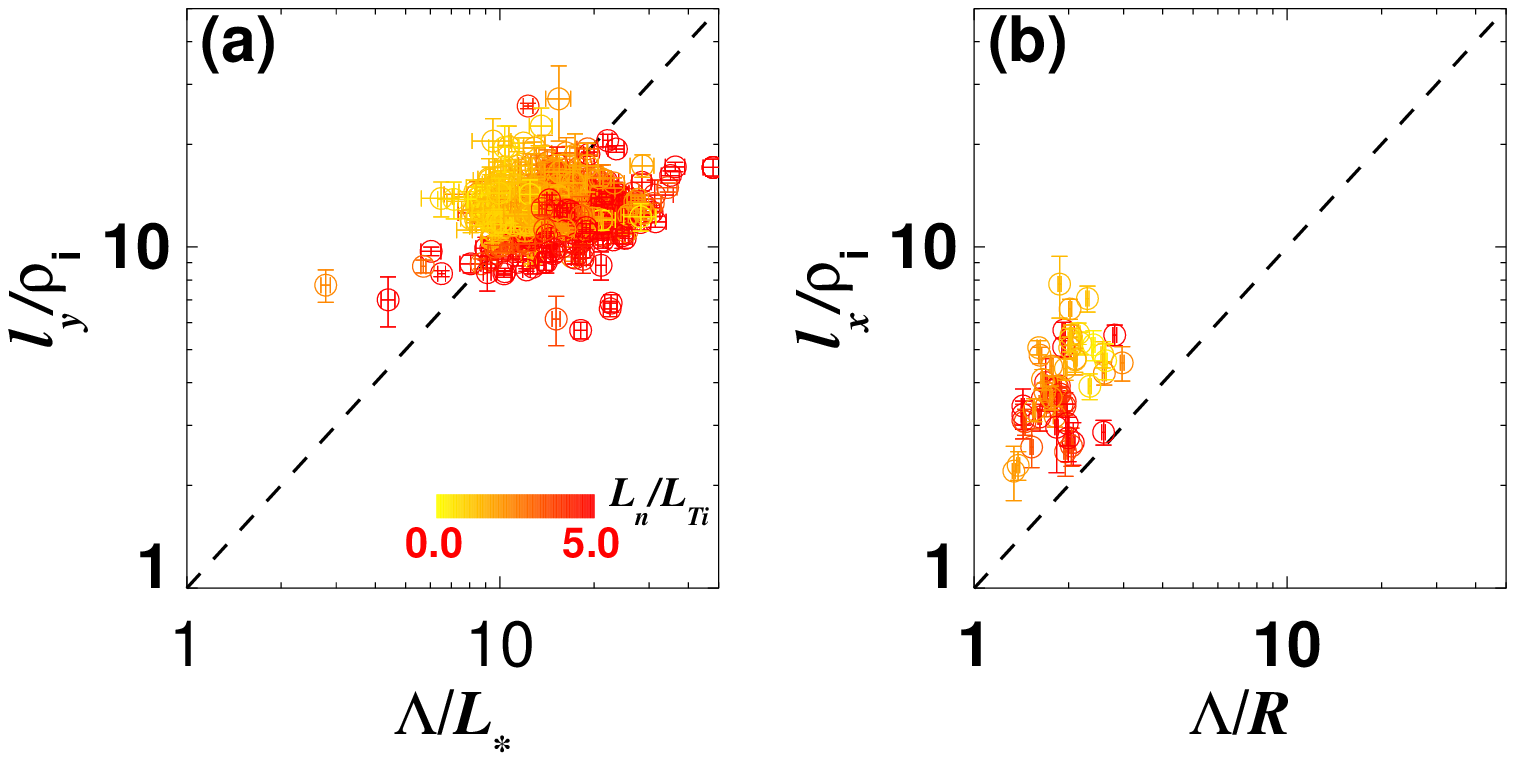}{(a) Poloidal correlation length $\ly/\rhoi$ vs.\ $\Lambda/\Lstar$; (b) Radial correlation length $\lx/\rhoi$ vs.\ $\Lambda/R$. Color as in \figref{all_vs_tau_c_eta}.}
This suggests that the balance of all relevant timescales determines correlation scales of the turbulence in all three spatial directions. Indeed, balancing $\tM\sim\tst$, we find the radial correlation scale $\lx/\rhoi\sim \Lambda/R$, the scaling tested in \figref{spatial_eta}(b), with a degree of success. This means that the density fluctuations we are measuring in MAST are not isotropic in the perpendicular plane, but rather elongated in the poloidal direction $\ly/\lx\sim R/\Lstar$ ($\sim5$ in our data). Interestingly, this clashes with the reported approximate isotropy ($\lx\sim\ly$) both in Cyclone Base Case simulations \cite{barnes_prl_2011_107} and in measured DIII-D turbulence (where $\ly/\lx\sim1.4$ \cite{shafer_pop_2012} and $\lx$ does not appear to depend on $B_p$ \cite{rhodes_pop_2002}). Whether this is a difference between spherical and conventional tokamaks is not as yet clear. 

\paragraph{Nonlinear time.}
Since we know the fluctuation amplitude, we can directly estimate the time scale associated with the advection of the fluctuations ($\vec{\delvper}\cdot\vec{\nabla}\dn$) by the fluctuating $\vec{E}\times\vec{B}$ velocity $\delvper = c\vec{B}\times\vec{\nabla}\varphi/B^2$. The electrostatic potential $\varphi$ is not directly measured, but can be estimated assuming Boltzmann response of the electrons: $\dn/n\approx e\varphi/T_e$. This estimate ignores trapped particles and, more importantly as we are about to argue, also does not apply to ion-scale zonal flows (poloidally and toroidally symmetric perturbations of $\varphi$ with $\dn=0$ \cite{diamond_ppcfreview_2005,fujisawa_nf_2009}). Thus, the non-zonal nonlinear time~is
\begin{equation}
\label{eq:tnl}
\lp\tnlnz\rp^{-1}=\frac{\vti\rhoi}{\lx\ly}\frac{T_e}{T_i}\frac{\dn}{n}.
\end{equation}
\myfig[3.0in]{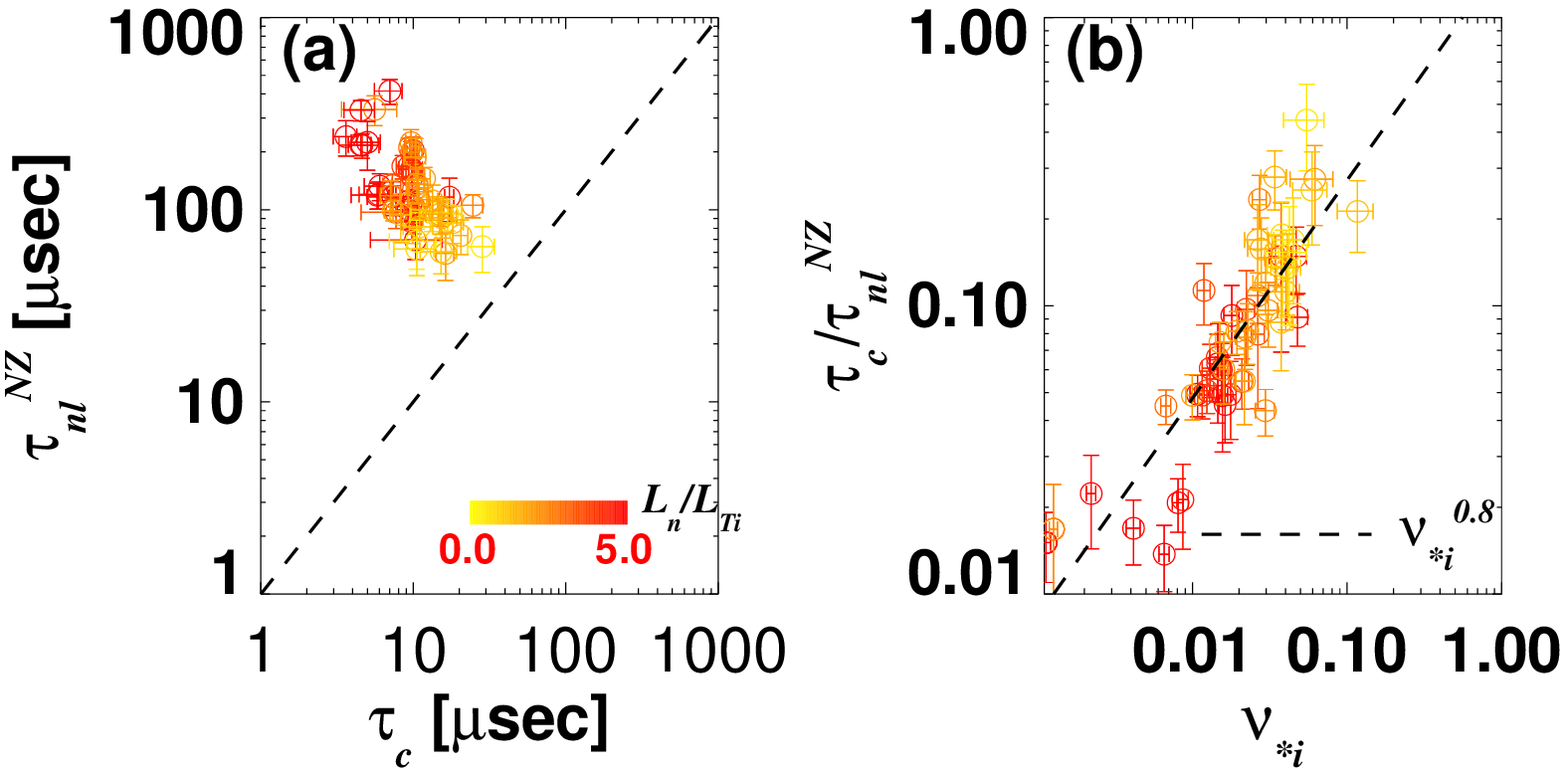}{(a) The nonlinear time associated with density fluctuations, $\tnlnz$, vs.\ the correlation time $\tc$; (b) their ratio vs.\ normalized ion collision rate $\nu_{*i}=\nu_{ii}\tst$. Color as in \figref{all_vs_tau_c_eta}.}
\figref{tau_nl_eta}(a) shows that $\tnlnz$ is always larger than $\tc$ (or the other time scales discussed above) and, furthermore, observed to have an inverse rather than direct correlation with it. Since turbulence clearly cannot be saturated by linear physics alone, this means that our estimate does not capture the correct nonlinear time. We conjecture that it is in fact the coupling to the zonal flows, invisible to BES (because their $\dn=0$), that dominates over the nonlinear interaction between the drift-wave-like fluctuations represented by $\tnlnz$ \cite{hammett_ppcf_1993, waltz_pop_1994, lin_science_1998, dimits_pop_2000, rogers_prl_2000, diamond_ppcfreview_2005, nakata_pop_2012, makwana_pop_2012}. It has long been suspected that the relative amplitude of the zonal flows compared to that of the drift waves depends on the ion collisionality \cite{hinton_ppcf_1999,diamond_ppcfreview_2005,xiao_pop_2007, ricci_prl_2006}. We can test this expectation by assuming that $\tc$ is the characteristic time associated with the coupling of the drift waves to the zonal flows and so depends on their amplitude. \figref{tau_nl_eta}(b) indeed shows a strong collisionality dependence: $\tc/\tnlnz \sim \nu_{*i}^{0.8\pm0.1}$, where $\nu_{*i}=\nu_{ii}\tst$ \footnote{A similar scaling is obtained for vs.\ $\nu_{ii}\tc$ and $\nu_{ii}\tstar$ or just straightforwardly for $(\tnlnz)^{-1}$ vs.\ $\nu_{ii}$.} (the ion collision time itself, $\nu_{ii}^{-1}$, is at least an order of magnitude longer than the time scales that participate in the ``grand critical balance''). If $\tc^{-1}\sim (\vti\rhoi/\lx\ly)e\varphi^{\rm ZF}/T_i$, where $\varphi^{\rm ZF}$ is the amplitude of the zonal potential, this result implies that the ratio of zonal to non-zonal component of the turbulence is $\varphi^{\rm ZF}/\varphi^{\rm NZ}\sim \nu_{*i}^{-0.8\pm0.1}$ \footnote{A scaling popular in theoretical models of zonal-flow-ITG turbulence is $\nu_{*i}^{-1/2}$\cite{diamond_ppcfreview_2005}.}.  
\newline\indent
We note that this situation is qualitatively distinct from what is seen in numerical simulations of ITG turbulence far from the threshold \cite{barnes_prl_2011_107}, where the drift-wave nonlinearity appears to dominate ($\tnlnz\sim\tc$). However, the turbulence in a real tokamak is likely to be close to marginal and so possibly in the state of reduced transport controlled by weakly-collisionally damped zonal flows \cite{fujisawa_nf_2009} and usually associated with the so-called ``Dimits upshift'' of the stiff-transport threshold \cite{dimits_pop_2000,lin_prl_1999,rogers_prl_2000,mikkelsen_prl_2008}. 

\paragraph{Discussion.} 
Our results support the notion that the statistics of turbulence are determined by the local equilibrium properties of the plasma. We find little correlation between the quantities reported above and the radial location \footnote{There is a slight bias in \figref{tau_nl_eta} for larger $\nu_{*i}$ to be found farther from the magnetic axis.} (note that we have limited our consideration to temporal and spatial scales and did not touch on the fluctuation amplitudes or transport properties, which do of course depend on radius). Our results also appeared insensitive to (i.e., not measurably correlated with) three other parameters that might in principle have proven important: $T_i/T_e$ (varied between $0.5$ and $2$), the magnetic shear $\hat s=d\ln q/d\ln r$ (varied between $-1$ and $5$) and the perpendicular component of the toroidal velocity shear $\tshear^{-1} = (B_p/B)dU_\phi/dr$. In much of our data, $\tshear \ge \tc,\tst$ (see \figref{all_vs_tau_c_eta}(d)), so it stands to reason that the statistics of the turbulence would not be dramatically affected; in the instances of $\tshear\sim\tst$, the effect of $\tshear$ could not be isolated \footnote{In general, we expect that a strong velocity shear would change $\lpar$ via a modified critical balance: if $\tshear<\tst$, then $\tc\sim\tshear\sim\lpar/\vti$, so $\lpar\sim\vti\tshear<\Lambda$.}. It would be interesting to investigate higher-rotation plasmas, as $\tshear^{-1}$, when sufficiently large, is expected to have a dramatic effect on transport \cite{waltz_pop_1994,devries_nf_2009,highcock_prl_2010,barnes_prl_2011,parra_prl_2011,mantica_prl_2011,highcock_prl_2012, roach_ppcf_2009}; even in our database, there is in fact some evidence that velocity shear might raise the critical temperature gradients \cite{ghim_prl2}, but we see no signature of this effect in the correlation properties of the turbulence.  

\paragraph{Conclusion.}
We have presented experimental results statistically consistent with a turbulent state in MAST set by the local equilibrium and in which the time scales of the linear drive, turbulence decorrelation, ion streaming and magnetic drifts are all similar and scale together as equilibrium parameters are varied. This ``grand critical balance'' implies a three-dimensionally anisotropic turbulence, with parallel, poloidal and radial correlation lengths  having different parameter dependences and $\lpar\gg\ly>\lx$. Our results also suggest the presence of a zonal component with an amplitude $\nu_{*i}^{-0.8\pm0.1}$ greater than the drift-wave density fluctuations.

\begin{acknowledgments}
We thank T.~Carter, J.~Connor, W.~Dorland, E.~Highcock, G.~McKee, C.~Michael, C.~Roach, J.~B.~Taylor and M.~Valovi\v{c} for valuable discussions. This work was supported in part by the RCUK Energy Programme under grant EP/I501045, the Kwanjeong Educational Foundation (Y-cG), the European Communities under the contract of Association between EURATOM and CCFE (Y-cG, ARF, IGA, GC) and by the Leverhulme Trust International Network for Magnetised Plasma Turbulence (MB, FIP). The views and opinions expressed herein do not necessarily reflect those of the European Commission.
\end{acknowledgments}


\bibliography{reference}

\begin{thebibliography}{67}%
\makeatletter
\providecommand \@ifxundefined [1]{%
 \@ifx{#1\undefined}
}%
\providecommand \@ifnum [1]{%
 \ifnum #1\expandafter \@firstoftwo
 \else \expandafter \@secondoftwo
 \fi
}%
\providecommand \@ifx [1]{%
 \ifx #1\expandafter \@firstoftwo
 \else \expandafter \@secondoftwo
 \fi
}%
\providecommand \natexlab [1]{#1}%
\providecommand \enquote  [1]{``#1''}%
\providecommand \bibnamefont  [1]{#1}%
\providecommand \bibfnamefont [1]{#1}%
\providecommand \citenamefont [1]{#1}%
\providecommand \href@noop [0]{\@secondoftwo}%
\providecommand \href [0]{\begingroup \@sanitize@url \@href}%
\providecommand \@href[1]{\@@startlink{#1}\@@href}%
\providecommand \@@href[1]{\endgroup#1\@@endlink}%
\providecommand \@sanitize@url [0]{\catcode `\\12\catcode `\$12\catcode
  `\&12\catcode `\#12\catcode `\^12\catcode `\_12\catcode `\%12\relax}%
\providecommand \@@startlink[1]{}%
\providecommand \@@endlink[0]{}%
\providecommand \url  [0]{\begingroup\@sanitize@url \@url }%
\providecommand \@url [1]{\endgroup\@href {#1}{\urlprefix }}%
\providecommand \urlprefix  [0]{URL }%
\providecommand \Eprint [0]{\href }%
\providecommand \doibase [0]{http://dx.doi.org/}%
\providecommand \selectlanguage [0]{\@gobble}%
\providecommand \bibinfo  [0]{\@secondoftwo}%
\providecommand \bibfield  [0]{\@secondoftwo}%
\providecommand \translation [1]{[#1]}%
\providecommand \BibitemOpen [0]{}%
\providecommand \bibitemStop [0]{}%
\providecommand \bibitemNoStop [0]{.\EOS\space}%
\providecommand \EOS [0]{\spacefactor3000\relax}%
\providecommand \BibitemShut  [1]{\csname bibitem#1\endcsname}%
\let\auto@bib@innerbib\@empty
\bibitem [{\citenamefont {Rudakov}\ and\ \citenamefont
  {Sagdeev}(1961)}]{rudakov_doklady_1961}%
  \BibitemOpen
  \bibfield  {author} {\bibinfo {author} {\bibfnamefont {L.~I.}\ \bibnamefont
  {Rudakov}}\ and\ \bibinfo {author} {\bibfnamefont {R.~Z.}\ \bibnamefont
  {Sagdeev}},\ }\href@noop {} {\bibfield  {journal} {\bibinfo  {journal} {Dokl.
  Akad. Nauk SSSR}\ }\textbf {\bibinfo {volume} {138}},\ \bibinfo {pages} {581}
  (\bibinfo {year} {1961})}\BibitemShut {NoStop}%
\bibitem [{\citenamefont {{Coppi}}\ \emph {et~al.}(1967)\citenamefont
  {{Coppi}}, \citenamefont {{Rosenbluth}},\ and\ \citenamefont
  {{Sagdeev}}}]{coppi_pof_1967}%
  \BibitemOpen
  \bibfield  {author} {\bibinfo {author} {\bibfnamefont {B.}~\bibnamefont
  {{Coppi}}}, \bibinfo {author} {\bibfnamefont {M.~N.}\ \bibnamefont
  {{Rosenbluth}}}, \ and\ \bibinfo {author} {\bibfnamefont {R.~Z.}\
  \bibnamefont {{Sagdeev}}},\ }\href {\doibase 10.1063/1.1762151} {\bibfield
  {journal} {\bibinfo  {journal} {Phys. Fluids}\ }\textbf {\bibinfo {volume}
  {10}},\ \bibinfo {pages} {582} (\bibinfo {year} {1967})}\BibitemShut
  {NoStop}%
\bibitem [{\citenamefont {{Cowley}}\ \emph {et~al.}(1991)\citenamefont
  {{Cowley}}, \citenamefont {{Kulsrud}},\ and\ \citenamefont
  {{Sudan}}}]{cowley_pfb_1991}%
  \BibitemOpen
  \bibfield  {author} {\bibinfo {author} {\bibfnamefont {S.~C.}\ \bibnamefont
  {{Cowley}}}, \bibinfo {author} {\bibfnamefont {R.~M.}\ \bibnamefont
  {{Kulsrud}}}, \ and\ \bibinfo {author} {\bibfnamefont {R.}~\bibnamefont
  {{Sudan}}},\ }\href {\doibase 10.1063/1.859913} {\bibfield  {journal}
  {\bibinfo  {journal} {Phys. Fluids B}\ }\textbf {\bibinfo {volume} {3}},\
  \bibinfo {pages} {2767} (\bibinfo {year} {1991})}\BibitemShut {NoStop}%
\bibitem [{\citenamefont {{Horton}}(1999)}]{horton_rmp_1999}%
  \BibitemOpen
  \bibfield  {author} {\bibinfo {author} {\bibfnamefont {W.}~\bibnamefont
  {{Horton}}},\ }\href {\doibase 10.1103/RevModPhys.71.735} {\bibfield
  {journal} {\bibinfo  {journal} {Rev. Mod. Phys.}\ }\textbf {\bibinfo {volume}
  {71}},\ \bibinfo {pages} {735} (\bibinfo {year} {1999})}\BibitemShut
  {NoStop}%
\bibitem [{\citenamefont {Barnes}\ \emph
  {et~al.}(2011{\natexlab{a}})\citenamefont {Barnes}, \citenamefont {Parra},\
  and\ \citenamefont {Schekochihin}}]{barnes_prl_2011_107}%
  \BibitemOpen
  \bibfield  {author} {\bibinfo {author} {\bibfnamefont {M.}~\bibnamefont
  {Barnes}}, \bibinfo {author} {\bibfnamefont {F.~I.}\ \bibnamefont {Parra}}, \
  and\ \bibinfo {author} {\bibfnamefont {A.~A.}\ \bibnamefont {Schekochihin}},\
  }\href@noop {} {\bibfield  {journal} {\bibinfo  {journal} {Phys. Rev. Lett.}\
  }\textbf {\bibinfo {volume} {107}},\ \bibinfo {pages} {115003} (\bibinfo
  {year} {2011}{\natexlab{a}})}\BibitemShut {NoStop}%
\bibitem [{\citenamefont {Goldreich}\ and\ \citenamefont
  {Sridhar}(1995)}]{goldreich_apj_1995}%
  \BibitemOpen
  \bibfield  {author} {\bibinfo {author} {\bibfnamefont {P.}~\bibnamefont
  {Goldreich}}\ and\ \bibinfo {author} {\bibfnamefont {S.}~\bibnamefont
  {Sridhar}},\ }\href@noop {} {\bibfield  {journal} {\bibinfo  {journal}
  {Astrophys. J.}\ }\textbf {\bibinfo {volume} {438}},\ \bibinfo {pages} {763}
  (\bibinfo {year} {1995})}\BibitemShut {NoStop}%
\bibitem [{\citenamefont {{Cho}}\ and\ \citenamefont
  {{Lazarian}}(2004)}]{cho_apj_2004}%
  \BibitemOpen
  \bibfield  {author} {\bibinfo {author} {\bibfnamefont {J.}~\bibnamefont
  {{Cho}}}\ and\ \bibinfo {author} {\bibfnamefont {A.}~\bibnamefont
  {{Lazarian}}},\ }\href {\doibase 10.1086/425215} {\bibfield  {journal}
  {\bibinfo  {journal} {Astrophys. J.}\ }\textbf {\bibinfo {volume} {615}},\
  \bibinfo {pages} {L41} (\bibinfo {year} {2004})}\BibitemShut {NoStop}%
\bibitem [{\citenamefont {{Schekochihin}}\ \emph {et~al.}(2009)\citenamefont
  {{Schekochihin}}, \citenamefont {{Cowley}}, \citenamefont {{Dorland}},
  \citenamefont {{Hammett}}, \citenamefont {{Howes}}, \citenamefont
  {{Quataert}},\ and\ \citenamefont {{Tatsuno}}}]{schekochihin_apjs_2009}%
  \BibitemOpen
  \bibfield  {author} {\bibinfo {author} {\bibfnamefont {A.~A.}\ \bibnamefont
  {{Schekochihin}}}, \bibinfo {author} {\bibfnamefont {S.~C.}\ \bibnamefont
  {{Cowley}}}, \bibinfo {author} {\bibfnamefont {W.}~\bibnamefont {{Dorland}}},
  \bibinfo {author} {\bibfnamefont {G.~W.}\ \bibnamefont {{Hammett}}}, \bibinfo
  {author} {\bibfnamefont {G.~G.}\ \bibnamefont {{Howes}}}, \bibinfo {author}
  {\bibfnamefont {E.}~\bibnamefont {{Quataert}}}, \ and\ \bibinfo {author}
  {\bibfnamefont {T.}~\bibnamefont {{Tatsuno}}},\ }\href {\doibase
  10.1088/0067-0049/182/1/310} {\bibfield  {journal} {\bibinfo  {journal}
  {Astrophys. J. Suppl.}\ }\textbf {\bibinfo {volume} {182}},\ \bibinfo {pages}
  {310} (\bibinfo {year} {2009})}\BibitemShut {NoStop}%
\bibitem [{\citenamefont {Nazarenko}\ and\ \citenamefont
  {Schekochihin}(2011)}]{nazarenko_jfm_2011}%
  \BibitemOpen
  \bibfield  {author} {\bibinfo {author} {\bibfnamefont {S.~V.}\ \bibnamefont
  {Nazarenko}}\ and\ \bibinfo {author} {\bibfnamefont {A.~A.}\ \bibnamefont
  {Schekochihin}},\ }\href@noop {} {\bibfield  {journal} {\bibinfo  {journal}
  {J. Fluid Mech.}\ }\textbf {\bibinfo {volume} {677}},\ \bibinfo {pages} {134}
  (\bibinfo {year} {2011})}\BibitemShut {NoStop}%
\bibitem [{\citenamefont {Horbury}\ \emph {et~al.}(2008)\citenamefont
  {Horbury}, \citenamefont {Forman},\ and\ \citenamefont
  {Oughton}}]{horbury_prl_2008}%
  \BibitemOpen
  \bibfield  {author} {\bibinfo {author} {\bibfnamefont {T.~S.}\ \bibnamefont
  {Horbury}}, \bibinfo {author} {\bibfnamefont {M.}~\bibnamefont {Forman}}, \
  and\ \bibinfo {author} {\bibfnamefont {S.}~\bibnamefont {Oughton}},\
  }\href@noop {} {\bibfield  {journal} {\bibinfo  {journal} {Phys. Rev. Lett.}\
  }\textbf {\bibinfo {volume} {101}},\ \bibinfo {pages} {175005} (\bibinfo
  {year} {2008})}\BibitemShut {NoStop}%
\bibitem [{\citenamefont {{Podesta}}(2009)}]{podesta_apj_2009}%
  \BibitemOpen
  \bibfield  {author} {\bibinfo {author} {\bibfnamefont {J.~J.}\ \bibnamefont
  {{Podesta}}},\ }\href {\doibase 10.1088/0004-637X/698/2/986} {\bibfield
  {journal} {\bibinfo  {journal} {Astrophys. J.}\ }\textbf {\bibinfo {volume}
  {698}},\ \bibinfo {pages} {986} (\bibinfo {year} {2009})}\BibitemShut
  {NoStop}%
\bibitem [{\citenamefont {{Wicks}}\ \emph {et~al.}(2010)\citenamefont
  {{Wicks}}, \citenamefont {{Horbury}}, \citenamefont {{Chen}},\ and\
  \citenamefont {{Schekochihin}}}]{wicks_mnras_2010}%
  \BibitemOpen
  \bibfield  {author} {\bibinfo {author} {\bibfnamefont {R.~T.}\ \bibnamefont
  {{Wicks}}}, \bibinfo {author} {\bibfnamefont {T.~S.}\ \bibnamefont
  {{Horbury}}}, \bibinfo {author} {\bibfnamefont {C.~H.~K.}\ \bibnamefont
  {{Chen}}}, \ and\ \bibinfo {author} {\bibfnamefont {A.~A.}\ \bibnamefont
  {{Schekochihin}}},\ }\href {\doibase 10.1111/j.1745-3933.2010.00898.x}
  {\bibfield  {journal} {\bibinfo  {journal} {Mon. Not. R. Astron. Soc.}\
  }\textbf {\bibinfo {volume} {407}},\ \bibinfo {pages} {L31} (\bibinfo {year}
  {2010})}\BibitemShut {NoStop}%
\bibitem [{\citenamefont {{Cho}}\ and\ \citenamefont
  {{Vishniac}}(2000)}]{cho_apj_2000}%
  \BibitemOpen
  \bibfield  {author} {\bibinfo {author} {\bibfnamefont {J.}~\bibnamefont
  {{Cho}}}\ and\ \bibinfo {author} {\bibfnamefont {E.~T.}\ \bibnamefont
  {{Vishniac}}},\ }\href {\doibase 10.1086/309213} {\bibfield  {journal}
  {\bibinfo  {journal} {Astrophys. J.}\ }\textbf {\bibinfo {volume} {539}},\
  \bibinfo {pages} {273} (\bibinfo {year} {2000})}\BibitemShut {NoStop}%
\bibitem [{\citenamefont {{Maron}}\ and\ \citenamefont
  {{Goldreich}}(2001)}]{maron_apj_2001}%
  \BibitemOpen
  \bibfield  {author} {\bibinfo {author} {\bibfnamefont {J.}~\bibnamefont
  {{Maron}}}\ and\ \bibinfo {author} {\bibfnamefont {P.}~\bibnamefont
  {{Goldreich}}},\ }\href {\doibase 10.1086/321413} {\bibfield  {journal}
  {\bibinfo  {journal} {Astrophys. J.}\ }\textbf {\bibinfo {volume} {554}},\
  \bibinfo {pages} {1175} (\bibinfo {year} {2001})}\BibitemShut {NoStop}%
\bibitem [{\citenamefont {{Chen}}\ \emph {et~al.}(2011)\citenamefont {{Chen}},
  \citenamefont {{Mallet}}, \citenamefont {{Yousef}}, \citenamefont
  {{Schekochihin}},\ and\ \citenamefont {{Horbury}}}]{chen_mnras_2011}%
  \BibitemOpen
  \bibfield  {author} {\bibinfo {author} {\bibfnamefont {C.~H.~K.}\
  \bibnamefont {{Chen}}}, \bibinfo {author} {\bibfnamefont {A.}~\bibnamefont
  {{Mallet}}}, \bibinfo {author} {\bibfnamefont {T.~A.}\ \bibnamefont
  {{Yousef}}}, \bibinfo {author} {\bibfnamefont {A.~A.}\ \bibnamefont
  {{Schekochihin}}}, \ and\ \bibinfo {author} {\bibfnamefont {T.~S.}\
  \bibnamefont {{Horbury}}},\ }\href {\doibase
  10.1111/j.1365-2966.2011.18933.x} {\bibfield  {journal} {\bibinfo  {journal}
  {Mon. Not. R. Astron. Soc.}\ }\textbf {\bibinfo {volume} {415}},\ \bibinfo
  {pages} {3219} (\bibinfo {year} {2011})}\BibitemShut {NoStop}%
\bibitem [{\citenamefont {TenBarge}\ and\ \citenamefont
  {Howes}(2012)}]{tenbarge_pop_2012}%
  \BibitemOpen
  \bibfield  {author} {\bibinfo {author} {\bibfnamefont {J.~M.}\ \bibnamefont
  {TenBarge}}\ and\ \bibinfo {author} {\bibfnamefont {G.~G.}\ \bibnamefont
  {Howes}},\ }\href@noop {} {\bibfield  {journal} {\bibinfo  {journal} {Phys.
  Plasmas}\ }\textbf {\bibinfo {volume} {19}},\ \bibinfo {pages} {055901}
  (\bibinfo {year} {2012})}\BibitemShut {NoStop}%
\bibitem [{\citenamefont {Fonck}\ \emph {et~al.}(1990)\citenamefont {Fonck},
  \citenamefont {Duperrex},\ and\ \citenamefont {Paul}}]{fonck_rsi_1990}%
  \BibitemOpen
  \bibfield  {author} {\bibinfo {author} {\bibfnamefont {R.~J.}\ \bibnamefont
  {Fonck}}, \bibinfo {author} {\bibfnamefont {P.~A.}\ \bibnamefont {Duperrex}},
  \ and\ \bibinfo {author} {\bibfnamefont {S.~F.}\ \bibnamefont {Paul}},\
  }\href@noop {} {\bibfield  {journal} {\bibinfo  {journal} {Rev. Sci.
  Instrum.}\ }\textbf {\bibinfo {volume} {61}},\ \bibinfo {pages} {3487}
  (\bibinfo {year} {1990})}\BibitemShut {NoStop}%
\bibitem [{\citenamefont {Fonck}\ \emph {et~al.}(1993)\citenamefont {Fonck},
  \citenamefont {Cosby}, \citenamefont {Durst}, \citenamefont {Paul},
  \citenamefont {Bretz}, \citenamefont {Scott}, \citenamefont {Synakowski},\
  and\ \citenamefont {Taylor}}]{fonck_prl_1993}%
  \BibitemOpen
  \bibfield  {author} {\bibinfo {author} {\bibfnamefont {R.~J.}\ \bibnamefont
  {Fonck}}, \bibinfo {author} {\bibfnamefont {G.}~\bibnamefont {Cosby}},
  \bibinfo {author} {\bibfnamefont {R.~D.}\ \bibnamefont {Durst}}, \bibinfo
  {author} {\bibfnamefont {S.~F.}\ \bibnamefont {Paul}}, \bibinfo {author}
  {\bibfnamefont {N.}~\bibnamefont {Bretz}}, \bibinfo {author} {\bibfnamefont
  {S.}~\bibnamefont {Scott}}, \bibinfo {author} {\bibfnamefont
  {E.}~\bibnamefont {Synakowski}}, \ and\ \bibinfo {author} {\bibfnamefont
  {G.}~\bibnamefont {Taylor}},\ }\href@noop {} {\bibfield  {journal} {\bibinfo
  {journal} {Phys. Rev. Lett.}\ }\textbf {\bibinfo {volume} {70}},\ \bibinfo
  {pages} {3736} (\bibinfo {year} {1993})}\BibitemShut {NoStop}%
\bibitem [{\citenamefont {{McKee}}\ \emph {et~al.}(1999)\citenamefont
  {{McKee}}, \citenamefont {{Ashley}}, \citenamefont {{Durst}}, \citenamefont
  {{Fonck}}, \citenamefont {{Jakubowski}}, \citenamefont {{Tritz}},
  \citenamefont {{Burrell}}, \citenamefont {{Greenfield}},\ and\ \citenamefont
  {{Robinson}}}]{mckee_rsi_1999}%
  \BibitemOpen
  \bibfield  {author} {\bibinfo {author} {\bibfnamefont {G.}~\bibnamefont
  {{McKee}}}, \bibinfo {author} {\bibfnamefont {R.}~\bibnamefont {{Ashley}}},
  \bibinfo {author} {\bibfnamefont {R.}~\bibnamefont {{Durst}}}, \bibinfo
  {author} {\bibfnamefont {R.}~\bibnamefont {{Fonck}}}, \bibinfo {author}
  {\bibfnamefont {M.}~\bibnamefont {{Jakubowski}}}, \bibinfo {author}
  {\bibfnamefont {K.}~\bibnamefont {{Tritz}}}, \bibinfo {author} {\bibfnamefont
  {K.}~\bibnamefont {{Burrell}}}, \bibinfo {author} {\bibfnamefont
  {C.}~\bibnamefont {{Greenfield}}}, \ and\ \bibinfo {author} {\bibfnamefont
  {J.}~\bibnamefont {{Robinson}}},\ }\href {\doibase 10.1063/1.1149416}
  {\bibfield  {journal} {\bibinfo  {journal} {Rev. Sci. Instrum.}\ }\textbf
  {\bibinfo {volume} {70}},\ \bibinfo {pages} {913} (\bibinfo {year}
  {1999})}\BibitemShut {NoStop}%
\bibitem [{\citenamefont {{McKee}}\ \emph {et~al.}(2003)\citenamefont
  {{McKee}}, \citenamefont {{Fenzi}}, \citenamefont {{Fonck}},\ and\
  \citenamefont {{Jakubowski}}}]{mckee_rsi_2003}%
  \BibitemOpen
  \bibfield  {author} {\bibinfo {author} {\bibfnamefont {G.~R.}\ \bibnamefont
  {{McKee}}}, \bibinfo {author} {\bibfnamefont {C.}~\bibnamefont {{Fenzi}}},
  \bibinfo {author} {\bibfnamefont {R.~J.}\ \bibnamefont {{Fonck}}}, \ and\
  \bibinfo {author} {\bibfnamefont {M.}~\bibnamefont {{Jakubowski}}},\ }\href
  {\doibase 10.1063/1.1535248} {\bibfield  {journal} {\bibinfo  {journal} {Rev.
  Sci. Instrum.}\ }\textbf {\bibinfo {volume} {74}},\ \bibinfo {pages} {2014}
  (\bibinfo {year} {2003})}\BibitemShut {NoStop}%
\bibitem [{\citenamefont {Field}\ \emph {et~al.}(2012)\citenamefont {Field},
  \citenamefont {Dunai}, \citenamefont {Gaffka}, \citenamefont {Ghim},
  \citenamefont {Kiss}, \citenamefont {Meszaros}, \citenamefont {Krizsanoczi},
  \citenamefont {Shibaev},\ and\ \citenamefont {Zoletnik}}]{field_rsi_2012}%
  \BibitemOpen
  \bibfield  {author} {\bibinfo {author} {\bibfnamefont {A.~R.}\ \bibnamefont
  {Field}}, \bibinfo {author} {\bibfnamefont {D.}~\bibnamefont {Dunai}},
  \bibinfo {author} {\bibfnamefont {R.}~\bibnamefont {Gaffka}}, \bibinfo
  {author} {\bibfnamefont {Y.-c.}\ \bibnamefont {Ghim}}, \bibinfo {author}
  {\bibfnamefont {I.}~\bibnamefont {Kiss}}, \bibinfo {author} {\bibfnamefont
  {B.}~\bibnamefont {Meszaros}}, \bibinfo {author} {\bibfnamefont
  {T.}~\bibnamefont {Krizsanoczi}}, \bibinfo {author} {\bibfnamefont
  {S.}~\bibnamefont {Shibaev}}, \ and\ \bibinfo {author} {\bibfnamefont
  {S.}~\bibnamefont {Zoletnik}},\ }\href@noop {} {\bibfield  {journal}
  {\bibinfo  {journal} {Rev. Sci. Instrum.}\ }\textbf {\bibinfo {volume}
  {83}},\ \bibinfo {pages} {013508} (\bibinfo {year} {2012})}\BibitemShut
  {NoStop}%
\bibitem [{\citenamefont {McKee}\ \emph {et~al.}(2001)\citenamefont {McKee},
  \citenamefont {Petty}, \citenamefont {Waltz}, \citenamefont {Fenzi},
  \citenamefont {Fonck}, \citenamefont {Kinsey}, \citenamefont {Luce},
  \citenamefont {Burrell}, \citenamefont {Baker}, \citenamefont {Doyle},
  \citenamefont {Garbet}, \citenamefont {Moyer}, \citenamefont {Rettig},
  \citenamefont {Rhodes}, \citenamefont {Ross}, \citenamefont {Staebler},
  \citenamefont {Sydora},\ and\ \citenamefont {Wade}}]{mckee_nf_2001}%
  \BibitemOpen
  \bibfield  {author} {\bibinfo {author} {\bibfnamefont {G.~R.}\ \bibnamefont
  {McKee}}, \bibinfo {author} {\bibfnamefont {C.~C.}\ \bibnamefont {Petty}},
  \bibinfo {author} {\bibfnamefont {R.~E.}\ \bibnamefont {Waltz}}, \bibinfo
  {author} {\bibfnamefont {C.}~\bibnamefont {Fenzi}}, \bibinfo {author}
  {\bibfnamefont {R.~J.}\ \bibnamefont {Fonck}}, \bibinfo {author}
  {\bibfnamefont {J.~E.}\ \bibnamefont {Kinsey}}, \bibinfo {author}
  {\bibfnamefont {T.~C.}\ \bibnamefont {Luce}}, \bibinfo {author}
  {\bibfnamefont {K.~H.}\ \bibnamefont {Burrell}}, \bibinfo {author}
  {\bibfnamefont {D.~R.}\ \bibnamefont {Baker}}, \bibinfo {author}
  {\bibfnamefont {E.~J.}\ \bibnamefont {Doyle}}, \bibinfo {author}
  {\bibfnamefont {X.}~\bibnamefont {Garbet}}, \bibinfo {author} {\bibfnamefont
  {R.~A.}\ \bibnamefont {Moyer}}, \bibinfo {author} {\bibfnamefont {C.~L.}\
  \bibnamefont {Rettig}}, \bibinfo {author} {\bibfnamefont {T.~L.}\
  \bibnamefont {Rhodes}}, \bibinfo {author} {\bibfnamefont {D.~W.}\
  \bibnamefont {Ross}}, \bibinfo {author} {\bibfnamefont {G.~M.}\ \bibnamefont
  {Staebler}}, \bibinfo {author} {\bibfnamefont {R.}~\bibnamefont {Sydora}}, \
  and\ \bibinfo {author} {\bibfnamefont {M.~R.}\ \bibnamefont {Wade}},\
  }\href@noop {} {\bibfield  {journal} {\bibinfo  {journal} {Nucl. Fusion}\
  }\textbf {\bibinfo {volume} {41}},\ \bibinfo {pages} {1235} (\bibinfo {year}
  {2001})}\BibitemShut {NoStop}%
\bibitem [{\citenamefont {Hennequin}\ \emph {et~al.}(2004)\citenamefont
  {Hennequin}, \citenamefont {Sabot}, \citenamefont {Honore}, \citenamefont
  {Hoang}, \citenamefont {Garbet}, \citenamefont {Truc}, \citenamefont
  {Fenzi},\ and\ \citenamefont {Quemeneur}}]{hennequin_ppcf_2004}%
  \BibitemOpen
  \bibfield  {author} {\bibinfo {author} {\bibfnamefont {P.}~\bibnamefont
  {Hennequin}}, \bibinfo {author} {\bibfnamefont {R.}~\bibnamefont {Sabot}},
  \bibinfo {author} {\bibfnamefont {C.}~\bibnamefont {Honore}}, \bibinfo
  {author} {\bibfnamefont {G.~T.}\ \bibnamefont {Hoang}}, \bibinfo {author}
  {\bibfnamefont {X.}~\bibnamefont {Garbet}}, \bibinfo {author} {\bibfnamefont
  {A.}~\bibnamefont {Truc}}, \bibinfo {author} {\bibfnamefont {C.}~\bibnamefont
  {Fenzi}}, \ and\ \bibinfo {author} {\bibfnamefont {A.}~\bibnamefont
  {Quemeneur}},\ }\href@noop {} {\bibfield  {journal} {\bibinfo  {journal}
  {Plasma Phys. Control. Fusion}\ }\textbf {\bibinfo {volume} {46}},\ \bibinfo
  {pages} {B121} (\bibinfo {year} {2004})}\BibitemShut {NoStop}%
\bibitem [{\citenamefont {Hutchinson}(2002)}]{hutchinson_ppcf_2002}%
  \BibitemOpen
  \bibfield  {author} {\bibinfo {author} {\bibfnamefont {I.~H.}\ \bibnamefont
  {Hutchinson}},\ }\href@noop {} {\bibfield  {journal} {\bibinfo  {journal}
  {Plasma Phys. Control. Fusion}\ }\textbf {\bibinfo {volume} {44}},\ \bibinfo
  {pages} {71} (\bibinfo {year} {2002})}\BibitemShut {NoStop}%
\bibitem [{\citenamefont {Scannell}\ \emph {et~al.}(2010)\citenamefont
  {Scannell}, \citenamefont {Walsh}, \citenamefont {Dunstan}, \citenamefont
  {Figueiredo}, \citenamefont {Naylor}, \citenamefont {O'Gorman}, \citenamefont
  {Shibaev}, \citenamefont {Gibson},\ and\ \citenamefont
  {Wilson}}]{scannell_rsi_2010}%
  \BibitemOpen
  \bibfield  {author} {\bibinfo {author} {\bibfnamefont {R.}~\bibnamefont
  {Scannell}}, \bibinfo {author} {\bibfnamefont {M.~J.}\ \bibnamefont {Walsh}},
  \bibinfo {author} {\bibfnamefont {M.~R.}\ \bibnamefont {Dunstan}}, \bibinfo
  {author} {\bibfnamefont {J.}~\bibnamefont {Figueiredo}}, \bibinfo {author}
  {\bibfnamefont {G.}~\bibnamefont {Naylor}}, \bibinfo {author} {\bibfnamefont
  {T.}~\bibnamefont {O'Gorman}}, \bibinfo {author} {\bibfnamefont
  {S.}~\bibnamefont {Shibaev}}, \bibinfo {author} {\bibfnamefont {K.~J.}\
  \bibnamefont {Gibson}}, \ and\ \bibinfo {author} {\bibfnamefont
  {H.}~\bibnamefont {Wilson}},\ }\href@noop {} {\bibfield  {journal} {\bibinfo
  {journal} {Rev. Sci. Instrum.}\ }\textbf {\bibinfo {volume} {81}},\ \bibinfo
  {pages} {10D520} (\bibinfo {year} {2010})}\BibitemShut {NoStop}%
\bibitem [{\citenamefont {Conway}\ \emph {et~al.}(2006)\citenamefont {Conway},
  \citenamefont {Carolan}, \citenamefont {McCone}, \citenamefont {Walsh},\ and\
  \citenamefont {Wisse}}]{conway_rsi_2006}%
  \BibitemOpen
  \bibfield  {author} {\bibinfo {author} {\bibfnamefont {N.~J.}\ \bibnamefont
  {Conway}}, \bibinfo {author} {\bibfnamefont {P.~G.}\ \bibnamefont {Carolan}},
  \bibinfo {author} {\bibfnamefont {J.}~\bibnamefont {McCone}}, \bibinfo
  {author} {\bibfnamefont {M.~J.}\ \bibnamefont {Walsh}}, \ and\ \bibinfo
  {author} {\bibfnamefont {M.}~\bibnamefont {Wisse}},\ }\href@noop {}
  {\bibfield  {journal} {\bibinfo  {journal} {Rev. Sci. Instrum.}\ }\textbf
  {\bibinfo {volume} {77}},\ \bibinfo {pages} {10F131} (\bibinfo {year}
  {2006})}\BibitemShut {NoStop}%
\bibitem [{\citenamefont {De~Bock}\ \emph {et~al.}(2008)\citenamefont
  {De~Bock}, \citenamefont {Conway}, \citenamefont {Walsh}, \citenamefont
  {Carolan},\ and\ \citenamefont {Hawkes}}]{debock_rsi_2008}%
  \BibitemOpen
  \bibfield  {author} {\bibinfo {author} {\bibfnamefont {M.~F.~M.}\
  \bibnamefont {De~Bock}}, \bibinfo {author} {\bibfnamefont {N.~J.}\
  \bibnamefont {Conway}}, \bibinfo {author} {\bibfnamefont {M.~J.}\
  \bibnamefont {Walsh}}, \bibinfo {author} {\bibfnamefont {P.~G.}\ \bibnamefont
  {Carolan}}, \ and\ \bibinfo {author} {\bibfnamefont {N.~C.}\ \bibnamefont
  {Hawkes}},\ }\href@noop {} {\bibfield  {journal} {\bibinfo  {journal} {Rev.
  Sci. Instrum.}\ }\textbf {\bibinfo {volume} {79}},\ \bibinfo {pages} {10F524}
  (\bibinfo {year} {2008})}\BibitemShut {NoStop}%
\bibitem [{\citenamefont {Lao}\ \emph {et~al.}(1985)\citenamefont {Lao},
  \citenamefont {St~John}, \citenamefont {Stambaugh}, \citenamefont {Kellman},\
  and\ \citenamefont {Pfeiffer}}]{lao_nf_1985}%
  \BibitemOpen
  \bibfield  {author} {\bibinfo {author} {\bibfnamefont {L.~L.}\ \bibnamefont
  {Lao}}, \bibinfo {author} {\bibfnamefont {H.}~\bibnamefont {St~John}},
  \bibinfo {author} {\bibfnamefont {R.~D.}\ \bibnamefont {Stambaugh}}, \bibinfo
  {author} {\bibfnamefont {A.~G.}\ \bibnamefont {Kellman}}, \ and\ \bibinfo
  {author} {\bibfnamefont {W.}~\bibnamefont {Pfeiffer}},\ }\href@noop {}
  {\bibfield  {journal} {\bibinfo  {journal} {Nucl. Fusion}\ }\textbf {\bibinfo
  {volume} {25}},\ \bibinfo {pages} {1611} (\bibinfo {year}
  {1985})}\BibitemShut {NoStop}%
\bibitem [{\citenamefont {Ghim}\ \emph
  {et~al.}(2012{\natexlab{a}})\citenamefont {Ghim}, \citenamefont {Field},
  \citenamefont {Duani}, \citenamefont {Zoletnik}, \citenamefont {Bardoczi},
  \citenamefont {Schekochihin},\ and\ \citenamefont {the
  MAST~Team}}]{ghim_ppcf_2012}%
  \BibitemOpen
  \bibfield  {author} {\bibinfo {author} {\bibfnamefont {Y.-c.}\ \bibnamefont
  {Ghim}}, \bibinfo {author} {\bibfnamefont {A.~R.}\ \bibnamefont {Field}},
  \bibinfo {author} {\bibfnamefont {D.}~\bibnamefont {Duani}}, \bibinfo
  {author} {\bibfnamefont {S.}~\bibnamefont {Zoletnik}}, \bibinfo {author}
  {\bibfnamefont {L.}~\bibnamefont {Bardoczi}}, \bibinfo {author}
  {\bibfnamefont {A.~A.}\ \bibnamefont {Schekochihin}}, \ and\ \bibinfo
  {author} {\bibnamefont {the MAST~Team}},\ }\href@noop {} {\bibfield
  {journal} {\bibinfo  {journal} {Plasma Phys. Control. Fusion}\ }\textbf
  {\bibinfo {volume} {54}},\ \bibinfo {pages} {095012} (\bibinfo {year}
  {2012}{\natexlab{a}})}\BibitemShut {NoStop}%
\bibitem [{\citenamefont {Durst}\ \emph {et~al.}(1992)\citenamefont {Durst},
  \citenamefont {Fonck}, \citenamefont {Cosby}, \citenamefont {Evensen},\ and\
  \citenamefont {Paul}}]{durst_rsi_1992}%
  \BibitemOpen
  \bibfield  {author} {\bibinfo {author} {\bibfnamefont {R.~D.}\ \bibnamefont
  {Durst}}, \bibinfo {author} {\bibfnamefont {R.~J.}\ \bibnamefont {Fonck}},
  \bibinfo {author} {\bibfnamefont {G.}~\bibnamefont {Cosby}}, \bibinfo
  {author} {\bibfnamefont {H.}~\bibnamefont {Evensen}}, \ and\ \bibinfo
  {author} {\bibfnamefont {S.~F.}\ \bibnamefont {Paul}},\ }\href@noop {}
  {\bibfield  {journal} {\bibinfo  {journal} {Rev. Sci. Instrum.}\ }\textbf
  {\bibinfo {volume} {63}},\ \bibinfo {pages} {4907} (\bibinfo {year}
  {1992})}\BibitemShut {NoStop}%
\bibitem [{\citenamefont {Markwardt}(2008)}]{markwardt_mpfit}%
  \BibitemOpen
  \bibfield  {author} {\bibinfo {author} {\bibfnamefont {C.~B.}\ \bibnamefont
  {Markwardt}},\ }\href@noop {} {\bibfield  {journal} {\bibinfo  {journal}
  {``Non-Linear Least Squares Fitting in IDL with MPFIT," in proc. Astronomical
  Data Analysis Software and Systems XVIII, Quebec, Canada, ASP Conference
  Series, Vol. 411, eds. D. Bohlender, P. Dowler and D. Durand (Astronomical
  Society of the Pacific: San Francisco), p. 251-254}\ } (\bibinfo {year}
  {2008})}\BibitemShut {NoStop}%
\bibitem [{Note1()}]{Note1}%
  \BibitemOpen
  \bibinfo {note} {However, for $\tau _{\protect \rm c}\lesssim 10\protect
  \tmspace +\medmuskip {.2222em}\mu $sec, $\tau _{\ast e}\sim \tau _{\ast i}$
  and for $\tau _{\protect \rm c}\gtrsim 10\protect \tmspace +\medmuskip
  {.2222em}\mu $sec, $\tau _{\ast n}\sim \tau _{\ast i}$, so we cannot rule out
  ion-scale electron drive (e.g., trapped electron modes \cite
  {kadomtsev_nf_1971} or microtearing \cite
  {roach_ppcf_2005,guttenfelder_pop_2012,doerk_pop_2012}).}\BibitemShut {Stop}%
\bibitem [{Note2()}]{Note2}%
  \BibitemOpen
  \bibinfo {note} {As noted above, our method for measuring $\tau _{\protect
  \rm c}$ would instead yield $\ell _\parallel /\protect \ensuremath
  {v_{{\protect \rm th}i}}$ if ${\protect \rm Ma}>\protect \qopname \relax
  o{cos}\alpha $, but that would require much stronger rotation (the smallest
  value in our database is $\protect \qopname \relax o{cos}\alpha \approx
  0.76$).}\BibitemShut {Stop}%
\bibitem [{Note3()}]{Note3}%
  \BibitemOpen
  \bibinfo {note} {In a conventional tokamak, $\Lambda \approx \pi q R$, where
  $q$ is the safety factor and $R$ major radius, but in a spherical tokamak,
  the local estimate we use is more appropriate}\BibitemShut {NoStop}%
\bibitem [{Note4()}]{Note4}%
  \BibitemOpen
  \bibinfo {note} {As $\tau _{\protect \rm M}$ contains $\ell _x$, there are 8
  times fewer data points here than in previous two figures, as explained
  above}\BibitemShut {NoStop}%
\bibitem [{\citenamefont {Shafer}\ \emph {et~al.}(2012)\citenamefont {Shafer},
  \citenamefont {Fonck}, \citenamefont {McKee}, \citenamefont {Holland},
  \citenamefont {White},\ and\ \citenamefont {Schlossberg}}]{shafer_pop_2012}%
  \BibitemOpen
  \bibfield  {author} {\bibinfo {author} {\bibfnamefont {M.~W.}\ \bibnamefont
  {Shafer}}, \bibinfo {author} {\bibfnamefont {R.~J.}\ \bibnamefont {Fonck}},
  \bibinfo {author} {\bibfnamefont {G.~R.}\ \bibnamefont {McKee}}, \bibinfo
  {author} {\bibfnamefont {C.}~\bibnamefont {Holland}}, \bibinfo {author}
  {\bibfnamefont {A.~E.}\ \bibnamefont {White}}, \ and\ \bibinfo {author}
  {\bibfnamefont {D.~J.}\ \bibnamefont {Schlossberg}},\ }\href@noop {}
  {\bibfield  {journal} {\bibinfo  {journal} {Phys. Plasmas}\ }\textbf
  {\bibinfo {volume} {19}},\ \bibinfo {pages} {032504} (\bibinfo {year}
  {2012})}\BibitemShut {NoStop}%
\bibitem [{\citenamefont {{Rhodes}}\ \emph {et~al.}(2002)\citenamefont
  {{Rhodes}}, \citenamefont {{Leboeuf}}, \citenamefont {{Sydora}},
  \citenamefont {{Groebner}}, \citenamefont {{Doyle}}, \citenamefont {{McKee}},
  \citenamefont {{Peebles}}, \citenamefont {{Rettig}}, \citenamefont {{Zeng}},\
  and\ \citenamefont {{Wang}}}]{rhodes_pop_2002}%
  \BibitemOpen
  \bibfield  {author} {\bibinfo {author} {\bibfnamefont {T.~L.}\ \bibnamefont
  {{Rhodes}}}, \bibinfo {author} {\bibfnamefont {J.-N.}\ \bibnamefont
  {{Leboeuf}}}, \bibinfo {author} {\bibfnamefont {R.~D.}\ \bibnamefont
  {{Sydora}}}, \bibinfo {author} {\bibfnamefont {R.~J.}\ \bibnamefont
  {{Groebner}}}, \bibinfo {author} {\bibfnamefont {E.~J.}\ \bibnamefont
  {{Doyle}}}, \bibinfo {author} {\bibfnamefont {G.~R.}\ \bibnamefont
  {{McKee}}}, \bibinfo {author} {\bibfnamefont {W.~A.}\ \bibnamefont
  {{Peebles}}}, \bibinfo {author} {\bibfnamefont {C.~L.}\ \bibnamefont
  {{Rettig}}}, \bibinfo {author} {\bibfnamefont {L.}~\bibnamefont {{Zeng}}}, \
  and\ \bibinfo {author} {\bibfnamefont {G.}~\bibnamefont {{Wang}}},\ }\href
  {\doibase 10.1063/1.1464544} {\bibfield  {journal} {\bibinfo  {journal}
  {Phys. Plasmas}\ }\textbf {\bibinfo {volume} {9}},\ \bibinfo {pages} {2141}
  (\bibinfo {year} {2002})}\BibitemShut {NoStop}%
\bibitem [{\citenamefont {Diamond}\ \emph {et~al.}(2005)\citenamefont
  {Diamond}, \citenamefont {Itoh}, \citenamefont {Itoh},\ and\ \citenamefont
  {Hahm}}]{diamond_ppcfreview_2005}%
  \BibitemOpen
  \bibfield  {author} {\bibinfo {author} {\bibfnamefont {P.~H.}\ \bibnamefont
  {Diamond}}, \bibinfo {author} {\bibfnamefont {S.-I.}\ \bibnamefont {Itoh}},
  \bibinfo {author} {\bibfnamefont {K.}~\bibnamefont {Itoh}}, \ and\ \bibinfo
  {author} {\bibfnamefont {T.-S.}\ \bibnamefont {Hahm}},\ }\href@noop {}
  {\bibfield  {journal} {\bibinfo  {journal} {Plasma Phys. Control. Fusion}\
  }\textbf {\bibinfo {volume} {47}},\ \bibinfo {pages} {R35} (\bibinfo {year}
  {2005})}\BibitemShut {NoStop}%
\bibitem [{\citenamefont {{Fujisawa}}(2009)}]{fujisawa_nf_2009}%
  \BibitemOpen
  \bibfield  {author} {\bibinfo {author} {\bibfnamefont {A.}~\bibnamefont
  {{Fujisawa}}},\ }\href {\doibase 10.1088/0029-5515/49/1/013001} {\bibfield
  {journal} {\bibinfo  {journal} {Nucl. Fusion}\ }\textbf {\bibinfo {volume}
  {49}},\ \bibinfo {pages} {013001} (\bibinfo {year} {2009})}\BibitemShut
  {NoStop}%
\bibitem [{\citenamefont {{Hammett}}\ \emph {et~al.}(1993)\citenamefont
  {{Hammett}}, \citenamefont {{Beer}}, \citenamefont {{Dorland}}, \citenamefont
  {{Cowley}},\ and\ \citenamefont {{Smith}}}]{hammett_ppcf_1993}%
  \BibitemOpen
  \bibfield  {author} {\bibinfo {author} {\bibfnamefont {G.~W.}\ \bibnamefont
  {{Hammett}}}, \bibinfo {author} {\bibfnamefont {M.~A.}\ \bibnamefont
  {{Beer}}}, \bibinfo {author} {\bibfnamefont {W.}~\bibnamefont {{Dorland}}},
  \bibinfo {author} {\bibfnamefont {S.~C.}\ \bibnamefont {{Cowley}}}, \ and\
  \bibinfo {author} {\bibfnamefont {S.~A.}\ \bibnamefont {{Smith}}},\ }\href
  {\doibase 10.1088/0741-3335/35/8/006} {\bibfield  {journal} {\bibinfo
  {journal} {Plasma Phys. and Control. Fusion}\ }\textbf {\bibinfo {volume}
  {35}},\ \bibinfo {pages} {973} (\bibinfo {year} {1993})}\BibitemShut
  {NoStop}%
\bibitem [{\citenamefont {Waltz}\ \emph {et~al.}(1994)\citenamefont {Waltz},
  \citenamefont {Kerbel},\ and\ \citenamefont {Milovich}}]{waltz_pop_1994}%
  \BibitemOpen
  \bibfield  {author} {\bibinfo {author} {\bibfnamefont {R.~E.}\ \bibnamefont
  {Waltz}}, \bibinfo {author} {\bibfnamefont {G.~D.}\ \bibnamefont {Kerbel}}, \
  and\ \bibinfo {author} {\bibfnamefont {J.}~\bibnamefont {Milovich}},\
  }\href@noop {} {\bibfield  {journal} {\bibinfo  {journal} {Phys. Plasmas}\
  }\textbf {\bibinfo {volume} {1}},\ \bibinfo {pages} {2229} (\bibinfo {year}
  {1994})}\BibitemShut {NoStop}%
\bibitem [{\citenamefont {{Lin}}\ \emph {et~al.}(1998)\citenamefont {{Lin}},
  \citenamefont {{Hahm}}, \citenamefont {{Lee}}, \citenamefont {{Tang}},\ and\
  \citenamefont {{White}}}]{lin_science_1998}%
  \BibitemOpen
  \bibfield  {author} {\bibinfo {author} {\bibfnamefont {Z.}~\bibnamefont
  {{Lin}}}, \bibinfo {author} {\bibfnamefont {T.~S.}\ \bibnamefont {{Hahm}}},
  \bibinfo {author} {\bibfnamefont {W.~W.}\ \bibnamefont {{Lee}}}, \bibinfo
  {author} {\bibfnamefont {W.~M.}\ \bibnamefont {{Tang}}}, \ and\ \bibinfo
  {author} {\bibfnamefont {R.~B.}\ \bibnamefont {{White}}},\ }\href@noop {}
  {\bibfield  {journal} {\bibinfo  {journal} {Science}\ }\textbf {\bibinfo
  {volume} {281}},\ \bibinfo {pages} {1835} (\bibinfo {year}
  {1998})}\BibitemShut {NoStop}%
\bibitem [{\citenamefont {{Dimits}}\ \emph {et~al.}(2000)\citenamefont
  {{Dimits}}, \citenamefont {{Bateman}}, \citenamefont {{Beer}}, \citenamefont
  {{Cohen}}, \citenamefont {{Dorland}}, \citenamefont {{Hammett}},
  \citenamefont {{Kim}}, \citenamefont {{Kinsey}}, \citenamefont
  {{Kotschenreuther}}, \citenamefont {{Kritz}}, \citenamefont {{Lao}},
  \citenamefont {{Mandrekas}}, \citenamefont {{Nevins}}, \citenamefont
  {{Parker}}, \citenamefont {{Redd}}, \citenamefont {{Shumaker}}, \citenamefont
  {{Sydora}},\ and\ \citenamefont {{Weiland}}}]{dimits_pop_2000}%
  \BibitemOpen
  \bibfield  {author} {\bibinfo {author} {\bibfnamefont {A.~M.}\ \bibnamefont
  {{Dimits}}}, \bibinfo {author} {\bibfnamefont {G.}~\bibnamefont {{Bateman}}},
  \bibinfo {author} {\bibfnamefont {M.~A.}\ \bibnamefont {{Beer}}}, \bibinfo
  {author} {\bibfnamefont {B.~I.}\ \bibnamefont {{Cohen}}}, \bibinfo {author}
  {\bibfnamefont {W.}~\bibnamefont {{Dorland}}}, \bibinfo {author}
  {\bibfnamefont {G.~W.}\ \bibnamefont {{Hammett}}}, \bibinfo {author}
  {\bibfnamefont {C.}~\bibnamefont {{Kim}}}, \bibinfo {author} {\bibfnamefont
  {J.~E.}\ \bibnamefont {{Kinsey}}}, \bibinfo {author} {\bibfnamefont
  {M.}~\bibnamefont {{Kotschenreuther}}}, \bibinfo {author} {\bibfnamefont
  {A.~H.}\ \bibnamefont {{Kritz}}}, \bibinfo {author} {\bibfnamefont {L.~L.}\
  \bibnamefont {{Lao}}}, \bibinfo {author} {\bibfnamefont {J.}~\bibnamefont
  {{Mandrekas}}}, \bibinfo {author} {\bibfnamefont {W.~M.}\ \bibnamefont
  {{Nevins}}}, \bibinfo {author} {\bibfnamefont {S.~E.}\ \bibnamefont
  {{Parker}}}, \bibinfo {author} {\bibfnamefont {A.~J.}\ \bibnamefont
  {{Redd}}}, \bibinfo {author} {\bibfnamefont {D.~E.}\ \bibnamefont
  {{Shumaker}}}, \bibinfo {author} {\bibfnamefont {R.}~\bibnamefont
  {{Sydora}}}, \ and\ \bibinfo {author} {\bibfnamefont {J.}~\bibnamefont
  {{Weiland}}},\ }\href {\doibase 10.1063/1.873896} {\bibfield  {journal}
  {\bibinfo  {journal} {Phys. Plasmas}\ }\textbf {\bibinfo {volume} {7}},\
  \bibinfo {pages} {969} (\bibinfo {year} {2000})}\BibitemShut {NoStop}%
\bibitem [{\citenamefont {{Rogers}}\ \emph {et~al.}(2000)\citenamefont
  {{Rogers}}, \citenamefont {{Dorland}},\ and\ \citenamefont
  {{Kotschenreuther}}}]{rogers_prl_2000}%
  \BibitemOpen
  \bibfield  {author} {\bibinfo {author} {\bibfnamefont {B.~N.}\ \bibnamefont
  {{Rogers}}}, \bibinfo {author} {\bibfnamefont {W.}~\bibnamefont {{Dorland}}},
  \ and\ \bibinfo {author} {\bibfnamefont {M.}~\bibnamefont
  {{Kotschenreuther}}},\ }\href {\doibase 10.1103/PhysRevLett.85.5336}
  {\bibfield  {journal} {\bibinfo  {journal} {Phys. Rev. Lett.}\ }\textbf
  {\bibinfo {volume} {85}},\ \bibinfo {pages} {5336} (\bibinfo {year}
  {2000})}\BibitemShut {NoStop}%
\bibitem [{\citenamefont {Nakata}\ \emph {et~al.}(2012)\citenamefont {Nakata},
  \citenamefont {Watanabe},\ and\ \citenamefont {Sugama}}]{nakata_pop_2012}%
  \BibitemOpen
  \bibfield  {author} {\bibinfo {author} {\bibfnamefont {M.}~\bibnamefont
  {Nakata}}, \bibinfo {author} {\bibfnamefont {T.~H.}\ \bibnamefont
  {Watanabe}}, \ and\ \bibinfo {author} {\bibfnamefont {H.}~\bibnamefont
  {Sugama}},\ }\href@noop {} {\bibfield  {journal} {\bibinfo  {journal} {Phys.
  Plasmas}\ }\textbf {\bibinfo {volume} {19}},\ \bibinfo {pages} {022303}
  (\bibinfo {year} {2012})}\BibitemShut {NoStop}%
\bibitem [{\citenamefont {Makwana}\ \emph {et~al.}(2012)\citenamefont
  {Makwana}, \citenamefont {Terry},\ and\ \citenamefont
  {Kim}}]{makwana_pop_2012}%
  \BibitemOpen
  \bibfield  {author} {\bibinfo {author} {\bibfnamefont {K.~D.}\ \bibnamefont
  {Makwana}}, \bibinfo {author} {\bibfnamefont {P.~W.}\ \bibnamefont {Terry}},
  \ and\ \bibinfo {author} {\bibfnamefont {J.~H.}\ \bibnamefont {Kim}},\
  }\href@noop {} {\bibfield  {journal} {\bibinfo  {journal} {Phys. Plasmas}\
  }\textbf {\bibinfo {volume} {19}},\ \bibinfo {pages} {062310} (\bibinfo
  {year} {2012})}\BibitemShut {NoStop}%
\bibitem [{\citenamefont {Hinton}\ and\ \citenamefont
  {Rosenbluth}(1999)}]{hinton_ppcf_1999}%
  \BibitemOpen
  \bibfield  {author} {\bibinfo {author} {\bibfnamefont {F.~L.}\ \bibnamefont
  {Hinton}}\ and\ \bibinfo {author} {\bibfnamefont {M.~N.}\ \bibnamefont
  {Rosenbluth}},\ }\href@noop {} {\bibfield  {journal} {\bibinfo  {journal}
  {Plasma Phys. Control. Fusion}\ }\textbf {\bibinfo {volume} {41}},\ \bibinfo
  {pages} {A653} (\bibinfo {year} {1999})}\BibitemShut {NoStop}%
\bibitem [{\citenamefont {{Xiao}}\ \emph {et~al.}(2007)\citenamefont {{Xiao}},
  \citenamefont {{Catto}},\ and\ \citenamefont {{Molvig}}}]{xiao_pop_2007}%
  \BibitemOpen
  \bibfield  {author} {\bibinfo {author} {\bibfnamefont {Y.}~\bibnamefont
  {{Xiao}}}, \bibinfo {author} {\bibfnamefont {P.~J.}\ \bibnamefont {{Catto}}},
  \ and\ \bibinfo {author} {\bibfnamefont {K.}~\bibnamefont {{Molvig}}},\
  }\href {\doibase 10.1063/1.2536297} {\bibfield  {journal} {\bibinfo
  {journal} {Phys. Plasmas}\ }\textbf {\bibinfo {volume} {14}},\ \bibinfo
  {pages} {032302} (\bibinfo {year} {2007})}\BibitemShut {NoStop}%
\bibitem [{\citenamefont {Ricci}\ \emph {et~al.}(2006)\citenamefont {Ricci},
  \citenamefont {Rogers},\ and\ \citenamefont {Dorland}}]{ricci_prl_2006}%
  \BibitemOpen
  \bibfield  {author} {\bibinfo {author} {\bibfnamefont {P.}~\bibnamefont
  {Ricci}}, \bibinfo {author} {\bibfnamefont {B.~N.}\ \bibnamefont {Rogers}}, \
  and\ \bibinfo {author} {\bibfnamefont {W.}~\bibnamefont {Dorland}},\
  }\href@noop {} {\bibfield  {journal} {\bibinfo  {journal} {Phys. Rev. Lett.}\
  }\textbf {\bibinfo {volume} {97}},\ \bibinfo {pages} {245001} (\bibinfo
  {year} {2006})}\BibitemShut {NoStop}%
\bibitem [{Note5()}]{Note5}%
  \BibitemOpen
  \bibinfo {note} {A similar scaling is obtained for vs.\ $\nu _{ii}\tau
  _{\protect \rm c}$ and $\nu _{ii}\tau _\ast $ or just straightforwardly for
  $(\tau _{\protect \rm nl}^{\protect \rm NZ})^{-1}$ vs.\ $\nu
  _{ii}$.}\BibitemShut {Stop}%
\bibitem [{Note6()}]{Note6}%
  \BibitemOpen
  \bibinfo {note} {A scaling popular in theoretical models of zonal-flow-ITG
  turbulence is $\nu _{*i}^{-1/2}$\cite {diamond_ppcfreview_2005}.}\BibitemShut
  {Stop}%
\bibitem [{\citenamefont {{Lin}}\ \emph {et~al.}(1999)\citenamefont {{Lin}},
  \citenamefont {{Hahm}}, \citenamefont {{Lee}}, \citenamefont {{Tang}},\ and\
  \citenamefont {{Diamond}}}]{lin_prl_1999}%
  \BibitemOpen
  \bibfield  {author} {\bibinfo {author} {\bibfnamefont {Z.}~\bibnamefont
  {{Lin}}}, \bibinfo {author} {\bibfnamefont {T.~S.}\ \bibnamefont {{Hahm}}},
  \bibinfo {author} {\bibfnamefont {W.~W.}\ \bibnamefont {{Lee}}}, \bibinfo
  {author} {\bibfnamefont {W.~M.}\ \bibnamefont {{Tang}}}, \ and\ \bibinfo
  {author} {\bibfnamefont {P.~H.}\ \bibnamefont {{Diamond}}},\ }\href {\doibase
  10.1103/PhysRevLett.83.3645} {\bibfield  {journal} {\bibinfo  {journal}
  {Phys. Rev. Lett.}\ }\textbf {\bibinfo {volume} {83}},\ \bibinfo {pages}
  {3645} (\bibinfo {year} {1999})}\BibitemShut {NoStop}%
\bibitem [{\citenamefont {{Mikkelsen}}\ and\ \citenamefont
  {{Dorland}}(2008)}]{mikkelsen_prl_2008}%
  \BibitemOpen
  \bibfield  {author} {\bibinfo {author} {\bibfnamefont {D.~R.}\ \bibnamefont
  {{Mikkelsen}}}\ and\ \bibinfo {author} {\bibfnamefont {W.}~\bibnamefont
  {{Dorland}}},\ }\href {\doibase 10.1103/PhysRevLett.101.135003} {\bibfield
  {journal} {\bibinfo  {journal} {Phys. Rev. Letters}\ }\textbf {\bibinfo
  {volume} {101}},\ \bibinfo {eid} {135003} (\bibinfo {year}
  {2008})}\BibitemShut {NoStop}%
\bibitem [{Note7()}]{Note7}%
  \BibitemOpen
  \bibinfo {note} {There is a slight bias in Fig.~\ref {fig:tau_nl_eta} for
  larger $\nu _{*i}$ to be found farther from the magnetic axis.}\BibitemShut
  {Stop}%
\bibitem [{Note8()}]{Note8}%
  \BibitemOpen
  \bibinfo {note} {In general, we expect that a strong velocity shear would
  change $\ell _\parallel $ via a modified critical balance: if $\tau
  _{\protect \rm sh}<\tau _{\protect \rm st}$, then $\tau _{\protect \rm c}\sim
  \tau _{\protect \rm sh}\sim \ell _\parallel /\protect \ensuremath
  {v_{{\protect \rm th}i}}$, so $\ell _\parallel \sim \protect \ensuremath
  {v_{{\protect \rm th}i}}\tau _{\protect \rm sh}<\Lambda $.}\BibitemShut
  {Stop}%
\bibitem [{\citenamefont {{de Vries}}\ \emph {et~al.}(2009)\citenamefont {{de
  Vries}}, \citenamefont {{Joffrin}}, \citenamefont {{Brix}}, \citenamefont
  {{Challis}}, \citenamefont {{Cromb{\'e}}}, \citenamefont {{Esposito}},
  \citenamefont {{Hawkes}}, \citenamefont {{Giroud}}, \citenamefont {{Hobirk}},
  \citenamefont {{L{\"o}nnroth}}, \citenamefont {{Mantica}}, \citenamefont
  {{Strintzi}}, \citenamefont {{Tala}}, \citenamefont {{Voitsekhovitch}},\ and\
  \citenamefont {{JET-EFDA Contributors to the Work
  Programme}}}]{devries_nf_2009}%
  \BibitemOpen
  \bibfield  {author} {\bibinfo {author} {\bibfnamefont {P.~C.}\ \bibnamefont
  {{de Vries}}}, \bibinfo {author} {\bibfnamefont {E.}~\bibnamefont
  {{Joffrin}}}, \bibinfo {author} {\bibfnamefont {M.}~\bibnamefont {{Brix}}},
  \bibinfo {author} {\bibfnamefont {C.~D.}\ \bibnamefont {{Challis}}}, \bibinfo
  {author} {\bibfnamefont {K.}~\bibnamefont {{Cromb{\'e}}}}, \bibinfo {author}
  {\bibfnamefont {B.}~\bibnamefont {{Esposito}}}, \bibinfo {author}
  {\bibfnamefont {N.~C.}\ \bibnamefont {{Hawkes}}}, \bibinfo {author}
  {\bibfnamefont {C.}~\bibnamefont {{Giroud}}}, \bibinfo {author}
  {\bibfnamefont {J.}~\bibnamefont {{Hobirk}}}, \bibinfo {author}
  {\bibfnamefont {J.}~\bibnamefont {{L{\"o}nnroth}}}, \bibinfo {author}
  {\bibfnamefont {P.}~\bibnamefont {{Mantica}}}, \bibinfo {author}
  {\bibfnamefont {D.}~\bibnamefont {{Strintzi}}}, \bibinfo {author}
  {\bibfnamefont {T.}~\bibnamefont {{Tala}}}, \bibinfo {author} {\bibfnamefont
  {I.}~\bibnamefont {{Voitsekhovitch}}}, \ and\ \bibinfo {author} {\bibnamefont
  {{JET-EFDA Contributors to the Work Programme}}},\ }\href {\doibase
  10.1088/0029-5515/49/7/075007} {\bibfield  {journal} {\bibinfo  {journal}
  {Nucl. Fusion}\ }\textbf {\bibinfo {volume} {49}},\ \bibinfo {pages} {075007}
  (\bibinfo {year} {2009})}\BibitemShut {NoStop}%
\bibitem [{\citenamefont {Highcock}\ \emph {et~al.}(2010)\citenamefont
  {Highcock}, \citenamefont {Barnes}, \citenamefont {Schekochihin},
  \citenamefont {Parra}, \citenamefont {Roach},\ and\ \citenamefont
  {Cowley}}]{highcock_prl_2010}%
  \BibitemOpen
  \bibfield  {author} {\bibinfo {author} {\bibfnamefont {E.~G.}\ \bibnamefont
  {Highcock}}, \bibinfo {author} {\bibfnamefont {M.}~\bibnamefont {Barnes}},
  \bibinfo {author} {\bibfnamefont {A.~A.}\ \bibnamefont {Schekochihin}},
  \bibinfo {author} {\bibfnamefont {F.~I.}\ \bibnamefont {Parra}}, \bibinfo
  {author} {\bibfnamefont {C.~M.}\ \bibnamefont {Roach}}, \ and\ \bibinfo
  {author} {\bibfnamefont {S.~C.}\ \bibnamefont {Cowley}},\ }\href@noop {}
  {\bibfield  {journal} {\bibinfo  {journal} {Phys. Rev. Lett.}\ }\textbf
  {\bibinfo {volume} {105}},\ \bibinfo {pages} {215003} (\bibinfo {year}
  {2010})}\BibitemShut {NoStop}%
\bibitem [{\citenamefont {Barnes}\ \emph
  {et~al.}(2011{\natexlab{b}})\citenamefont {Barnes}, \citenamefont {Parra},
  \citenamefont {Highcock}, \citenamefont {Schekochihin}, \citenamefont
  {Cowley},\ and\ \citenamefont {Roach}}]{barnes_prl_2011}%
  \BibitemOpen
  \bibfield  {author} {\bibinfo {author} {\bibfnamefont {M.}~\bibnamefont
  {Barnes}}, \bibinfo {author} {\bibfnamefont {F.~I.}\ \bibnamefont {Parra}},
  \bibinfo {author} {\bibfnamefont {E.~G.}\ \bibnamefont {Highcock}}, \bibinfo
  {author} {\bibfnamefont {A.~A.}\ \bibnamefont {Schekochihin}}, \bibinfo
  {author} {\bibfnamefont {S.~C.}\ \bibnamefont {Cowley}}, \ and\ \bibinfo
  {author} {\bibfnamefont {C.~M.}\ \bibnamefont {Roach}},\ }\href@noop {}
  {\bibfield  {journal} {\bibinfo  {journal} {Phys. Rev. Lett.}\ }\textbf
  {\bibinfo {volume} {106}},\ \bibinfo {pages} {175004} (\bibinfo {year}
  {2011}{\natexlab{b}})}\BibitemShut {NoStop}%
\bibitem [{\citenamefont {{Parra}}\ \emph {et~al.}(2011)\citenamefont
  {{Parra}}, \citenamefont {{Barnes}}, \citenamefont {{Highcock}},
  \citenamefont {{Schekochihin}},\ and\ \citenamefont
  {{Cowley}}}]{parra_prl_2011}%
  \BibitemOpen
  \bibfield  {author} {\bibinfo {author} {\bibfnamefont {F.~I.}\ \bibnamefont
  {{Parra}}}, \bibinfo {author} {\bibfnamefont {M.}~\bibnamefont {{Barnes}}},
  \bibinfo {author} {\bibfnamefont {E.~G.}\ \bibnamefont {{Highcock}}},
  \bibinfo {author} {\bibfnamefont {A.~A.}\ \bibnamefont {{Schekochihin}}}, \
  and\ \bibinfo {author} {\bibfnamefont {S.~C.}\ \bibnamefont {{Cowley}}},\
  }\href {\doibase 10.1103/PhysRevLett.106.115004} {\bibfield  {journal}
  {\bibinfo  {journal} {Phys. Rev. Lett.}\ }\textbf {\bibinfo {volume} {106}},\
  \bibinfo {eid} {115004} (\bibinfo {year} {2011})}\BibitemShut {NoStop}%
\bibitem [{\citenamefont {Mantica}\ \emph {et~al.}(2011)\citenamefont
  {Mantica}, \citenamefont {Angioni}, \citenamefont {Challis}, \citenamefont
  {Colyer}, \citenamefont {Frassinetti}, \citenamefont {Hawkes}, \citenamefont
  {Johnson}, \citenamefont {Tsalas}, \citenamefont {deVries}, \citenamefont
  {Weiland}, \citenamefont {Baiocchi}, \citenamefont {Beurskens}, \citenamefont
  {Figueiredo}, \citenamefont {Giroud}, \citenamefont {Hobirk}, \citenamefont
  {Joffrin}, \citenamefont {Lerche}, \citenamefont {Naulin}, \citenamefont
  {Peeters}, \citenamefont {Salmi}, \citenamefont {Sozzi}, \citenamefont
  {Strintzi}, \citenamefont {Staebler}, \citenamefont {Tala}, \citenamefont
  {Van~Eester},\ and\ \citenamefont {Versloot}}]{mantica_prl_2011}%
  \BibitemOpen
  \bibfield  {author} {\bibinfo {author} {\bibfnamefont {P.}~\bibnamefont
  {Mantica}}, \bibinfo {author} {\bibfnamefont {C.}~\bibnamefont {Angioni}},
  \bibinfo {author} {\bibfnamefont {C.}~\bibnamefont {Challis}}, \bibinfo
  {author} {\bibfnamefont {G.}~\bibnamefont {Colyer}}, \bibinfo {author}
  {\bibfnamefont {L.}~\bibnamefont {Frassinetti}}, \bibinfo {author}
  {\bibfnamefont {N.}~\bibnamefont {Hawkes}}, \bibinfo {author} {\bibfnamefont
  {T.}~\bibnamefont {Johnson}}, \bibinfo {author} {\bibfnamefont
  {M.}~\bibnamefont {Tsalas}}, \bibinfo {author} {\bibfnamefont {P.~C.}\
  \bibnamefont {deVries}}, \bibinfo {author} {\bibfnamefont {J.}~\bibnamefont
  {Weiland}}, \bibinfo {author} {\bibfnamefont {B.}~\bibnamefont {Baiocchi}},
  \bibinfo {author} {\bibfnamefont {M.~N.~A.}\ \bibnamefont {Beurskens}},
  \bibinfo {author} {\bibfnamefont {A.~C.~A.}\ \bibnamefont {Figueiredo}},
  \bibinfo {author} {\bibfnamefont {C.}~\bibnamefont {Giroud}}, \bibinfo
  {author} {\bibfnamefont {J.}~\bibnamefont {Hobirk}}, \bibinfo {author}
  {\bibfnamefont {E.}~\bibnamefont {Joffrin}}, \bibinfo {author} {\bibfnamefont
  {E.}~\bibnamefont {Lerche}}, \bibinfo {author} {\bibfnamefont
  {V.}~\bibnamefont {Naulin}}, \bibinfo {author} {\bibfnamefont {A.~G.}\
  \bibnamefont {Peeters}}, \bibinfo {author} {\bibfnamefont {A.}~\bibnamefont
  {Salmi}}, \bibinfo {author} {\bibfnamefont {C.}~\bibnamefont {Sozzi}},
  \bibinfo {author} {\bibfnamefont {D.}~\bibnamefont {Strintzi}}, \bibinfo
  {author} {\bibfnamefont {G.}~\bibnamefont {Staebler}}, \bibinfo {author}
  {\bibfnamefont {T.}~\bibnamefont {Tala}}, \bibinfo {author} {\bibfnamefont
  {D.}~\bibnamefont {Van~Eester}}, \ and\ \bibinfo {author} {\bibfnamefont
  {T.}~\bibnamefont {Versloot}},\ }\href@noop {} {\bibfield  {journal}
  {\bibinfo  {journal} {Phys. Rev. Lett.}\ }\textbf {\bibinfo {volume} {107}},\
  \bibinfo {pages} {135004} (\bibinfo {year} {2011})}\BibitemShut {NoStop}%
\bibitem [{\citenamefont {{Highcock}}\ \emph {et~al.}(2012)\citenamefont
  {{Highcock}}, \citenamefont {{Schekochihin}}, \citenamefont {{Cowley}},
  \citenamefont {{Barnes}}, \citenamefont {{Parra}}, \citenamefont {{Roach}},\
  and\ \citenamefont {{Dorland}}}]{highcock_prl_2012}%
  \BibitemOpen
  \bibfield  {author} {\bibinfo {author} {\bibfnamefont {E.~G.}\ \bibnamefont
  {{Highcock}}}, \bibinfo {author} {\bibfnamefont {A.~A.}\ \bibnamefont
  {{Schekochihin}}}, \bibinfo {author} {\bibfnamefont {S.~C.}\ \bibnamefont
  {{Cowley}}}, \bibinfo {author} {\bibfnamefont {M.}~\bibnamefont {{Barnes}}},
  \bibinfo {author} {\bibfnamefont {F.~I.}\ \bibnamefont {{Parra}}}, \bibinfo
  {author} {\bibfnamefont {C.~M.}\ \bibnamefont {{Roach}}}, \ and\ \bibinfo
  {author} {\bibfnamefont {W.}~\bibnamefont {{Dorland}}},\ }\href@noop {}
  {\bibfield  {journal} {\bibinfo  {journal} {arXiv:1203.6455}\ } (\bibinfo
  {year} {2012})}\BibitemShut {NoStop}%
\bibitem [{\citenamefont {Roach}\ \emph {et~al.}(2009)\citenamefont {Roach},
  \citenamefont {Abel}, \citenamefont {Akers}, \citenamefont {Arter},
  \citenamefont {Barnes}, \citenamefont {Camenen}, \citenamefont {Casson},
  \citenamefont {Colyer}, \citenamefont {Connor}, \citenamefont {Cowley},
  \citenamefont {Dickinson}, \citenamefont {Dorland}, \citenamefont {Field},
  \citenamefont {Guttenfelder}, \citenamefont {Hammett}, \citenamefont
  {Hastie}, \citenamefont {Highcock}, \citenamefont {Loureiro}, \citenamefont
  {Peeters}, \citenamefont {Reshko}, \citenamefont {Saarelma}, \citenamefont
  {Schekochihin}, \citenamefont {Valovic},\ and\ \citenamefont
  {Wilson}}]{roach_ppcf_2009}%
  \BibitemOpen
  \bibfield  {author} {\bibinfo {author} {\bibfnamefont {C.~M.}\ \bibnamefont
  {Roach}}, \bibinfo {author} {\bibfnamefont {I.~G.}\ \bibnamefont {Abel}},
  \bibinfo {author} {\bibfnamefont {R.~J.}\ \bibnamefont {Akers}}, \bibinfo
  {author} {\bibfnamefont {W.}~\bibnamefont {Arter}}, \bibinfo {author}
  {\bibfnamefont {M.}~\bibnamefont {Barnes}}, \bibinfo {author} {\bibfnamefont
  {Y.}~\bibnamefont {Camenen}}, \bibinfo {author} {\bibfnamefont {F.~J.}\
  \bibnamefont {Casson}}, \bibinfo {author} {\bibfnamefont {G.}~\bibnamefont
  {Colyer}}, \bibinfo {author} {\bibfnamefont {J.~W.}\ \bibnamefont {Connor}},
  \bibinfo {author} {\bibfnamefont {S.~C.}\ \bibnamefont {Cowley}}, \bibinfo
  {author} {\bibfnamefont {D.}~\bibnamefont {Dickinson}}, \bibinfo {author}
  {\bibfnamefont {W.}~\bibnamefont {Dorland}}, \bibinfo {author} {\bibfnamefont
  {A.~R.}\ \bibnamefont {Field}}, \bibinfo {author} {\bibfnamefont
  {W.}~\bibnamefont {Guttenfelder}}, \bibinfo {author} {\bibfnamefont {G.~W.}\
  \bibnamefont {Hammett}}, \bibinfo {author} {\bibfnamefont {R.~J.}\
  \bibnamefont {Hastie}}, \bibinfo {author} {\bibfnamefont {E.}~\bibnamefont
  {Highcock}}, \bibinfo {author} {\bibfnamefont {N.~F.}\ \bibnamefont
  {Loureiro}}, \bibinfo {author} {\bibfnamefont {A.~G.}\ \bibnamefont
  {Peeters}}, \bibinfo {author} {\bibfnamefont {M.}~\bibnamefont {Reshko}},
  \bibinfo {author} {\bibfnamefont {S.}~\bibnamefont {Saarelma}}, \bibinfo
  {author} {\bibfnamefont {A.~A.}\ \bibnamefont {Schekochihin}}, \bibinfo
  {author} {\bibfnamefont {M.}~\bibnamefont {Valovic}}, \ and\ \bibinfo
  {author} {\bibfnamefont {H.~R.}\ \bibnamefont {Wilson}},\ }\href@noop {}
  {\bibfield  {journal} {\bibinfo  {journal} {Plasma Phys. Control. Fusion}\
  }\textbf {\bibinfo {volume} {51}},\ \bibinfo {pages} {124020} (\bibinfo
  {year} {2009})}\BibitemShut {NoStop}%
\bibitem [{\citenamefont {Ghim}\ \emph
  {et~al.}(2012{\natexlab{b}})\citenamefont {Ghim} \emph {et~al.}}]{ghim_prl2}%
  \BibitemOpen
  \bibfield  {author} {\bibinfo {author} {\bibfnamefont {Y.-c.}\ \bibnamefont
  {Ghim}} \emph {et~al.},\ }\href@noop {} {\bibfield  {journal} {\bibinfo
  {journal} {in preparation}\ } (\bibinfo {year}
  {2012}{\natexlab{b}})}\BibitemShut {NoStop}%
\bibitem [{\citenamefont {Kadomtsev}\ and\ \citenamefont
  {Pogutse}(1971)}]{kadomtsev_nf_1971}%
  \BibitemOpen
  \bibfield  {author} {\bibinfo {author} {\bibfnamefont {B.~B.}\ \bibnamefont
  {Kadomtsev}}\ and\ \bibinfo {author} {\bibfnamefont {O.~P.}\ \bibnamefont
  {Pogutse}},\ }\href@noop {} {\bibfield  {journal} {\bibinfo  {journal} {Nucl.
  Fusion}\ }\textbf {\bibinfo {volume} {11}},\ \bibinfo {pages} {67} (\bibinfo
  {year} {1971})}\BibitemShut {NoStop}%
\bibitem [{\citenamefont {{Roach}}\ \emph {et~al.}(2005)\citenamefont
  {{Roach}}, \citenamefont {{Applegate}}, \citenamefont {{Connor}},
  \citenamefont {{Cowley}}, \citenamefont {{Dorland}}, \citenamefont
  {{Hastie}}, \citenamefont {{Joiner}}, \citenamefont {{Saarelma}},
  \citenamefont {{Schekochihin}}, \citenamefont {{Akers}}, \citenamefont
  {{Brickley}}, \citenamefont {{Field}}, \citenamefont {{Valovic}},\ and\
  \citenamefont {{MAST Team}}}]{roach_ppcf_2005}%
  \BibitemOpen
  \bibfield  {author} {\bibinfo {author} {\bibfnamefont {C.~M.}\ \bibnamefont
  {{Roach}}}, \bibinfo {author} {\bibfnamefont {D.~J.}\ \bibnamefont
  {{Applegate}}}, \bibinfo {author} {\bibfnamefont {J.~W.}\ \bibnamefont
  {{Connor}}}, \bibinfo {author} {\bibfnamefont {S.~C.}\ \bibnamefont
  {{Cowley}}}, \bibinfo {author} {\bibfnamefont {W.~D.}\ \bibnamefont
  {{Dorland}}}, \bibinfo {author} {\bibfnamefont {R.~J.}\ \bibnamefont
  {{Hastie}}}, \bibinfo {author} {\bibfnamefont {N.}~\bibnamefont {{Joiner}}},
  \bibinfo {author} {\bibfnamefont {S.}~\bibnamefont {{Saarelma}}}, \bibinfo
  {author} {\bibfnamefont {A.~A.}\ \bibnamefont {{Schekochihin}}}, \bibinfo
  {author} {\bibfnamefont {R.~J.}\ \bibnamefont {{Akers}}}, \bibinfo {author}
  {\bibfnamefont {C.}~\bibnamefont {{Brickley}}}, \bibinfo {author}
  {\bibfnamefont {A.~R.}\ \bibnamefont {{Field}}}, \bibinfo {author}
  {\bibfnamefont {M.}~\bibnamefont {{Valovic}}}, \ and\ \bibinfo {author}
  {\bibnamefont {{MAST Team}}},\ }\href {\doibase 10.1088/0741-3335/47/12B/S23}
  {\bibfield  {journal} {\bibinfo  {journal} {Plasma Phys. Control. Fusion}\
  }\textbf {\bibinfo {volume} {47}},\ \bibinfo {pages} {B323} (\bibinfo {year}
  {2005})}\BibitemShut {NoStop}%
\bibitem [{\citenamefont {{Guttenfelder}}\ \emph {et~al.}(2012)\citenamefont
  {{Guttenfelder}}, \citenamefont {{Candy}}, \citenamefont {{Kaye}},
  \citenamefont {{Nevins}}, \citenamefont {{Wang}}, \citenamefont {{Zhang}},
  \citenamefont {{Bell}}, \citenamefont {{Crocker}}, \citenamefont {{Hammett}},
  \citenamefont {{LeBlanc}}, \citenamefont {{Mikkelsen}}, \citenamefont
  {{Ren}},\ and\ \citenamefont {{Yuh}}}]{guttenfelder_pop_2012}%
  \BibitemOpen
  \bibfield  {author} {\bibinfo {author} {\bibfnamefont {W.}~\bibnamefont
  {{Guttenfelder}}}, \bibinfo {author} {\bibfnamefont {J.}~\bibnamefont
  {{Candy}}}, \bibinfo {author} {\bibfnamefont {S.~M.}\ \bibnamefont {{Kaye}}},
  \bibinfo {author} {\bibfnamefont {W.~M.}\ \bibnamefont {{Nevins}}}, \bibinfo
  {author} {\bibfnamefont {E.}~\bibnamefont {{Wang}}}, \bibinfo {author}
  {\bibfnamefont {J.}~\bibnamefont {{Zhang}}}, \bibinfo {author} {\bibfnamefont
  {R.~E.}\ \bibnamefont {{Bell}}}, \bibinfo {author} {\bibfnamefont {N.~A.}\
  \bibnamefont {{Crocker}}}, \bibinfo {author} {\bibfnamefont {G.~W.}\
  \bibnamefont {{Hammett}}}, \bibinfo {author} {\bibfnamefont {B.~P.}\
  \bibnamefont {{LeBlanc}}}, \bibinfo {author} {\bibfnamefont {D.~R.}\
  \bibnamefont {{Mikkelsen}}}, \bibinfo {author} {\bibfnamefont
  {Y.}~\bibnamefont {{Ren}}}, \ and\ \bibinfo {author} {\bibfnamefont
  {H.}~\bibnamefont {{Yuh}}},\ }\href {\doibase 10.1063/1.3694104} {\bibfield
  {journal} {\bibinfo  {journal} {Phys. Plasmas}\ }\textbf {\bibinfo {volume}
  {19}},\ \bibinfo {pages} {056119} (\bibinfo {year} {2012})}\BibitemShut
  {NoStop}%
\bibitem [{\citenamefont {{Doerk}}\ \emph {et~al.}(2012)\citenamefont
  {{Doerk}}, \citenamefont {{Jenko}}, \citenamefont {{G{\"o}rler}},
  \citenamefont {{Told}}, \citenamefont {{Pueschel}},\ and\ \citenamefont
  {{Hatch}}}]{doerk_pop_2012}%
  \BibitemOpen
  \bibfield  {author} {\bibinfo {author} {\bibfnamefont {H.}~\bibnamefont
  {{Doerk}}}, \bibinfo {author} {\bibfnamefont {F.}~\bibnamefont {{Jenko}}},
  \bibinfo {author} {\bibfnamefont {T.}~\bibnamefont {{G{\"o}rler}}}, \bibinfo
  {author} {\bibfnamefont {D.}~\bibnamefont {{Told}}}, \bibinfo {author}
  {\bibfnamefont {M.~J.}\ \bibnamefont {{Pueschel}}}, \ and\ \bibinfo {author}
  {\bibfnamefont {D.~R.}\ \bibnamefont {{Hatch}}},\ }\href {\doibase
  10.1063/1.3694663} {\bibfield  {journal} {\bibinfo  {journal} {Phys.
  Plasmas}\ }\textbf {\bibinfo {volume} {19}},\ \bibinfo {pages} {055907}
  (\bibinfo {year} {2012})}\BibitemShut {NoStop}%
\end{thebibliography}%

\end{document}